\documentclass[aps,prl,reprint,amsmath,amssymb,showpacs,superscriptaddress]{revtex4-1}
\usepackage{CJK}
\usepackage{graphicx}
\usepackage{dcolumn}
\usepackage{bm}
\usepackage{color}
\usepackage{hyperref}

\begin{document}

\title{Reconciling light nuclei and nuclear matter: relativistic \textit{ab initio} calculations  }
\author{Y. L. Yang}
\affiliation{State Key Laboratory of Nuclear Physics and Technology, School of Physics, Peking University, Beijing 100871, China}

\author{P. W. Zhao}
\email{pwzhao@pku.edu.cn}
\affiliation{State Key Laboratory of Nuclear Physics and Technology, School of Physics, Peking University, Beijing 100871, China}

\begin{abstract}
An accurate and simultaneous \textit{ab initio} prediction for both light nuclei and nuclear matter has been a long-standing challenge in nuclear physics, due to the significant uncertainties associated with the three-nucleon forces. 
In this Letter, we develop the relativistic quantum Monte Carlo methods for the nuclear \textit{ab initio} problem, and
  calculate the ground-state energies of $A\leq4$ nuclei using the two-nucleon Bonn force with an unprecedented high accuracy. 
The present relativistic results significantly outperforms the nonrelativistic results with only two-nucleon forces. 
We demonstrate that both light nuclei and nuclear matter can be well described simultaneously in the relativistic \textit{ab initio} calculations, even in the absence of three-nucleon forces, and a correlation between the properties of light $A\leq4$ nuclei and the nuclear saturation is revealed.  
This provides a quantitative understanding of the connection between the light nuclei and nuclear matter saturation properties.
\end{abstract}

\maketitle

{\it Introduction.}---A major goal of nuclear theory is to explain the structure and properties of atomic nuclei in an \textit{ab initio} approach, where the nuclei are solved by nuclear many-body method as individual nucleons interacting via nuclear forces fixed in free-space scattering.
Already in the 1990s, various realistic models of two-nucleon ($NN$) forces had been constructed by fitting the $NN$ scattering observables to high precision.
Well-known and still widely used examples include the Argonne $v_{18}$ (AV18)~\cite{Wiringa1995Phys.Rev.C38} and Bonn~\cite{Machleidt1989,Machleidt2001Phys.Rev.C024001} potentials.
Despite their high-quality description of $NN$ scattering, these forces do not provide sufficient binding energies for $^3$H and $^4$He~\cite{Pudliner1995Phys.Rev.Lett.43964399,Navratil2000Phys.Rev.C054311}.
This leads to the discussion on the missing of a three-nucleon ($3N$) force, which is now widely  accepted in nuclear \textit{ab initio} calculations.

In the past two decades, many models of $3N$ forces have been  constructed~\cite{Epelbaum2002Phys.Rev.C064001,Navratil2007Phys.Rev.Lett.042501,Pieper2008AIPConf.Proc.143152,Gazit2009Phys.Rev.Lett.102502,
Lynn2016Phys.Rev.Lett.062501,Piarulli2018Phys.Rev.Lett.052503,Epelbaum2019Phys.Rev.C024313}, typically by adjusting to $A=3,4$ binding energies and/or other few-nucleon observables.
These $3N$ forces were then tested by calculating the bulk properties of heavier systems, among the most important ones being the saturation energy and density for infinite nuclear matter.
However, the predicted saturation properties varied greatly depending on the specific model and fitting details of the $3N$ forces~\cite{Coraggio2014Phys.Rev.C044321,Sammarruca2015Phys.Rev.C054311,Lonardoni2020Phys.Rev.Res.022033} (for a recent review, see Ref.~\cite{Hebeler2021Phys.Rep.1116}).
Consequently, a simultaneous reproduction of the properties of light nuclei and nuclear saturation becomes a challenging problem in nuclear \textit{ab initio} calculations~\cite{Hebeler2021Phys.Rep.1116}.
This has led some groups to fit nuclear forces, besides the $A=3,4$ binding energies, to properties of medium-mass nuclei, e.g., in Ref.~\cite{Ekstroem2015Phys.Rev.C051301}, or directly to the saturation properties of nuclear matter~\cite{Drischler2019Phys.Rev.Lett.042501}.

Given the aforementioned background, it is crucial to investigate any potential aspects that may have been omitted in the current \textit{ab initio} calculations.
One of such aspects is relativity, as most \textit{ab initio} calculations are currently performed within the nonrelativistic framework.
In fact, the typical momentum of the nucleons in nuclei is roughly 200 MeV, which is about 20\% of the nucleon mass, so the relativistic correction is expected to be around $p^2/M^2\simeq 4\%$.
If this accounts for 4\% of the binding energy, it is comparable to the contributions of $3N$ forces to the binding energies of $A=3,4$ nuclei.
If it accounts for 4\% of the kinetic or potential energies, then the relativistic effects may be even more significant, especially for nuclear saturation that results from delicate cancellations of  large potential and kinetic energies.

The significance of the relativistic effects has been recognized in the saturation of nuclear matter with the relativistic Brueckner-Hartree-Fock (RBHF) calculations \cite{Anastasio1980Phys.Rev.Lett.20962099,Brockmann1990Phys.Rev.C19651980,Sammarruca2012Phys.Rev.C054317,Wang2021Phys.Rev.C054319}.
In contrast to the nonrelativistic results, the RBHF calculations can provide a satisfactory description of the saturation energy and density, even without introducing a $3N$ force.
Similar calculations also provide reasonable results for the binding energies of several medium-mass nuclei with only $NN$ forces~\cite{Shen2016Chin.Phys.Lett.102103,Shen2017Phys.Rev.C014316,Shen2019Prog.Part.Nucl.Phys.103713}.

This raises a natural question of whether $A=3, 4$ nuclei and nuclear matter saturation properties can be simultaneously described in the relativistic framework.
While the RBHF theory is a good approximation for nuclear matter, it is less accurate for light nuclei due to the approximate treatment of the center-of-mass motions~\cite{Shen2017Phys.Rev.C014316}.
The existing relativistic three-body calculations~\cite{Stadler1997Phys.Rev.Lett.2629,Sammarruca1992Phys.Rev.C16361641} are specifically designed for $^3$H using different scattering equations from the RBHF one and, thus, cannot match the nuclear matter results from the RBHF calculations.
Moreover, their extensions to $A=4$ or heavier systems are formidable.
Therefore, a new relativistic \textit{ab initio} method is highly demanded.

Among the variety of \textit{ab initio} methods, the continuum Quantum Monte Carlo (QMC) methods~\cite{Carlson2015Rev.Mod.Phys.1067,Lynn2019Ann.Rev.Nucl.Part.Sci.279305,Gandolfi2020Front.Phys.} are notable for their high accuracy across various systems.
The variational Monte Carlo and Green's function Monte Carlo (GFMC) can provide reliable solutions for nuclei up to $A=12$~\cite{Pudliner1995Phys.Rev.Lett.43964399,Wiringa2002Phys.Rev.Lett.182501,Lovato2013Phys.Rev.Lett.092501,Piarulli2018Phys.Rev.Lett.052503}.
The auxiliary-field diffusion Monte Carlo (AFDMC) additionally samples spin-isospin degrees of freedom, making possible the study of larger nuclei and nuclear matter
~\cite{Gandolfi2007Phys.Rev.Lett.022507, Gandolfi2007Phys.Rev.Lett.102503, Lonardoni2018Phys.Rev.Lett.122502, Lonardoni2020Phys.Rev.Res.022033}.
Both the GFMC and AFDMC methods are based on the diffusion Monte Carlo algorithm that realizes the imaginary-time propagation stochastically and, thus, virtually exact.
However, the existing QMC methods are only successful for nearly local forces in the coordinate space~\cite{Carlson2015Rev.Mod.Phys.1067} and cannot be implemented in relativistic calculations, due to the nonlocality dictated by the covariance of nuclear forces.

In this Letter, we develop the relativistic QMC methods, including the relativistic variational Monte Carlo (RVMC) and relativistic diffusion Monte Carlo (RDMC),
and the ground states of $A\leq4$ nuclei are calculated using the Bonn forces with an unprecedentedly high accuracy.
A quantitative understanding of the connection between the light nuclei and nuclear matter saturation properties is provided, and the crucial role of the relativistic effects is demonstrated.

{\it The Bonn force.}---
The starting point is a covariant Lagrangian based on the meson-exchange picture~\cite{Machleidt1989},
\begin{equation}
  \begin{split}
  &\mathcal{L}_{\rm int}^{(s)}=g_s\overline{\psi}\psi\varphi^{(s)},\quad \mathcal{L}_{\rm int}^{(pv)}=-\frac{f_{ps}}{m_{ps}}\overline{\psi}\gamma^5\gamma^\mu\psi\partial_\mu\varphi^{(ps)}, \\
  &\mathcal{L}_{\rm int}^{(v)}=-g_v\overline{\psi}\gamma^\mu\psi\varphi_\mu^{(v)}
  -\frac{f_v}{4M}\overline{\psi}\sigma^{\mu\nu}\psi(\partial_\mu\varphi_\nu^{(v)}-\partial_\nu\varphi_\mu^{(v)}),
  \end{split}
\end{equation}
which includes the scalar (s) ($\sigma,\delta$), the pseudovector (pv) ($\eta,\pi$), and the vector (v) ($\omega,\rho$) meson-nucleon couplings, and in each channel it contains one isocalar meson and one isovector meson.
The Bonn force is then defined as the sum of one-meson-exchange amplitudes in the momentum space
\begin{equation}\label{Eq.bonn}
  \begin{split}
  V(\bm p,\bm p')=\frac{M^2}{E_{p'}E_{p}}\sum_{\varphi}&\overline{u}(\bm p')\Gamma_\phi u(\bm p)\overline{u}(-\bm p')\Gamma_\phi u(-\bm p)\\
  &\times\frac{[F_\phi(\bm p'-\bm p)]^2}{m_\phi^2+(\bm p'-\bm p)^2}
  \end{split}
\end{equation}
with $\bm p(\bm p')$ the initial (final) relative momentum of two nucleons, $M$ the nucleon mass, and $E_p=(p^2+M^2)^{1/2}$.
Here, $u$ is the positive-energy Dirac spinor, $\Gamma_\phi$ the interaction vertices in the Lagrangian with $F_\phi$ the corresponding form factors, and $m_\phi$ the physical meson masses.
The coupling strengths $g_\phi$ and the cutoff parameters in $F_\phi$ are determined by fitting $NN$ scattering data and deuteron properties with the relativistic Thompson equation.
Three parameter sets are available, denoted as $A$, $B$, and $C$~\cite{Machleidt1989}.

{\it The relativistic QMC methods.}---
For the ground states of nuclei, we solve the following many-body equation in the coordinate space $\{\bm R=(\bm r_1,\ldots,\bm r_A)\}$
\begin{equation}
  \hat{H}\Psi(\bm R)=\left(\sum_{i=1}^A \hat{K}_i+\sum_{i<j}^A \hat{V}_{ij}\right)\Psi(\bm R)=E\Psi(\bm R),
\end{equation}
including the relativistic kinetic part $\hat{K}_i$ and the $NN$ interaction part $\hat{V}_{ij}$,
\begin{equation}\label{Eq.K&V}
  \begin{split}
  \hat{K}_i\Psi(\bm R)&=\frac{M^2}{2\pi^2}\int{\rm d^3}r_i'\frac{K_2(M r_i')}{r_i'{}^2}[\Psi(\bm R)-\Psi(\bm R+\bm r_i')],\\
  \hat{V}_{ij}\Psi(\bm R)&=\int{\rm d^3}r_{ij}'V(\bm r_{ij},\bm r_{ij}')\Psi(\bm R+\bm r_{ij}'-\bm r_{ij}).
  \end{split}
\end{equation}
Here, $K_2$ is the modified Bessel function of order 2~\cite{Carlson1993Phys.Rev.C484} and $V(\bm r_{ij},\bm r_{ij}')$ is the coordinate-space representation of the Bonn potential with $\bm r_{ij}=\bm r_i-\bm r_j$.
The Coulomb force between finite-size protons are also considered~\cite{Wiringa1995Phys.Rev.C38}.
The present Hamiltonian differs from the nonrelativistic ones solved in the previous QMC methods, due to the strong nonlocality of the relativistic kinetic energy and the $NN$ interactions in the coordinate space.

Within the RVMC method, the ground state is solved by minimizing the energy expectation for a given form of trial function~$\Psi_T$
\begin{equation}
  E[\Psi_T]=\frac{\int{\rm d}\bm R \Psi_T^\dagger(\bm R)\hat{H}\Psi_T(\bm R)}{\int{\rm d}\bm R \Psi_T^\dagger(\bm R)\Psi_T(\bm R)}.
\end{equation}
The integral is evaluated with the standard Metropolis Monte Carlo sampling~\cite{Metropolis1953TheJournalofChemicalPhysics10871092}.
Different from the case with local interactions, the nonlocal integrals in Eq.~(\ref{Eq.K&V}) have to be evaluated for each sampled $\bm R$.
This is carried out with a combination of the stochastic spherical designs for the angular part~\cite{Fahy1988Phys.Rev.Lett.1631,Fahy1990Phys.Rev.B3503} and the Gaussian quadrature for the radial part.

The trial wave function is built with the neural-network ansatz~\cite{Adams2021Phys.Rev.Lett.022502,Yang2022Phys.Lett.B137587,Yang2023Phys.Rev.C034320},
\begin{equation}
  \Psi_T(\bm R)=\prod_{i<j}f_c(r_{ij})\left[
  1 + \sum_{i<j}\sum_{p=2}^6 u^p(r_{ij})O_{ij}^p
  \right]\Phi,
\end{equation}
where the pair correlation functions $f_c$ and $u^{p=2\text{-}6}$ in the Jastrow-like correlator are represented by feed-forward neural networks and $O^p_{ij}$ are two-body spin-isospin operators
\begin{equation}
  O_{ij}^{p=2\text{-}6}=\bm\tau_i\cdot\bm\tau_j,\bm\sigma_i\cdot\bm\sigma_j,\bm\sigma_i\cdot\bm\sigma_j\bm\tau_i\cdot\bm\tau_j,
  S_{ij}, S_{ij}\bm\tau_i\cdot\bm\tau_j,
\end{equation}
with $S_{ij}=3\bm\sigma\cdot\hat{\bm r}_{ij}\bm\sigma\cdot\hat{\bm r}_{ij}-\bm\sigma_i\cdot\bm\sigma_j$.
The Slater determinant $\Phi$ is taken to be $\mathcal{A}(\uparrow_n \uparrow_p)$ for $^2$H, $\mathcal{A}(\uparrow_n\downarrow_n\uparrow_p)$ for $^3$H, and $\mathcal{A}(\uparrow_n\downarrow_n\uparrow_p\downarrow_p)$ for $^4$He, with $\mathcal{A}$ being the antisymmetrization operator.
The neural networks contain variational parameters determined by minimizing the energy expectation iteratively.

\begin{figure}[!htpb]
    \centering
	\includegraphics[width=0.45\textwidth]{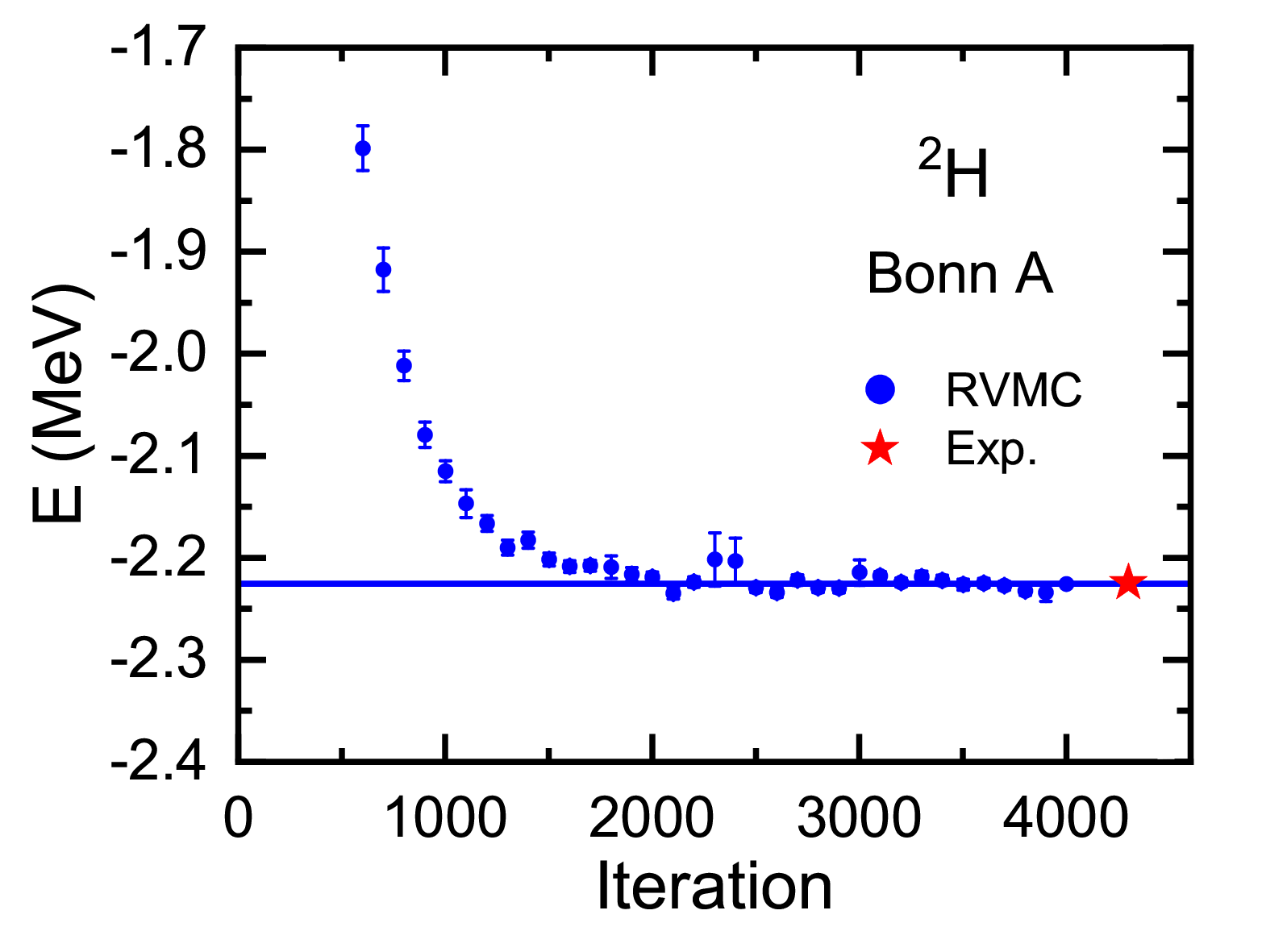}
	\caption{(Color online). The energy of $^2$H as a function of the iteration in the RVMC calculation with Bonn A.
The converged energy (solid line) is compared to the experimental value (star).
	}\label{fig1}
\end{figure}

Figure~\ref{fig1} depicts the energy expectation of $^2$H as a function of the iteration in the RVMC calculation with Bonn A.
The converged results perfectly reproduce the experimental energy of $^2$H within 1 keV.
This is not surprising since the $^2$H energy was used in the fit of the Bonn A potential~\cite{Machleidt1989} and the present trial wave function is, in principle, exact for two-body systems.
For $^3$H and $^4$He, however, the RVMC calculations provide slightly underbound solutions, due to the missing three- and four- body correlations in the trial wave function.
The missing energy can be evaluated by a subsequent RDMC calculation.

The RDMC method projects out the ground state $|\Psi_0\rangle$ via imaginary-time propagation starting from the trial wave function $|\Psi_T\rangle$,
\begin{equation}
  |\Psi_0\rangle\propto\lim_{\tau\rightarrow\infty}{\rm e}^{-\hat{H}\tau}|\Psi_T\rangle.
\end{equation}
It is realized by a sequence of short-time propagators $G^{\Delta\tau}_{\alpha\beta}(\bm R',\bm R)=\langle \bm R'\alpha|{\rm e}^{-\hat{H}\Delta\tau}|\bm R\beta\rangle$, where $\bm R,\bm R'$ are the coordinates before and after the propagation step, and $\alpha,\beta$ stand for spin-isospin states.
The short-time propagator can be split into the relativistic free propagator $G^{\Delta\tau}_{\alpha\beta}$ and the interaction term $I^{\Delta\tau}_{\alpha\beta}$~\cite{Carlson2015Rev.Mod.Phys.1067},
\begin{equation}\label{Eq.prop}
  \begin{split}
    G^{\Delta\tau}_{\alpha\beta}(\bm R',\bm R)&=G^{\Delta\tau}_0(\bm R',\bm R)I^{\Delta\tau}_{\alpha\beta}(\bm R',\bm R),
  \end{split}
\end{equation}
where
\begin{equation}\label{Eq.prop0}
  \begin{split}
    G^{\Delta\tau}_0(\bm R',\bm R)=&\prod_{i=1}^A\frac{M^2}{2\pi^2}\frac{\Delta\tau{\rm e}^{M\Delta\tau}}{(\bm r_i'-\bm r_i)^2+\Delta\tau^2}\\
    &\times K_2\left(M\sqrt{(\bm r_i'-\bm r_i)^2+\Delta\tau^2}\right).
  \end{split}
\end{equation}

To evaluate the interaction term $I^{\Delta\tau}_{\alpha\beta}(\bm R',\bm R)$, we define a local part of the $NN$ interaction, $\bar{V}(\bm r)=\int{\rm d}^3 \bm r' V(\bm r,\bm r')$, which dominates the potential energy in the RVMC calculations (about $95\%$).
The nonlocality in the propagator is considered effectively by replacing $\hat{V}-\bar{V}$ with $a\bar{V}$, where $a$ is fixed by $\langle \hat{V}-\bar{V}\rangle\simeq \langle a\bar{V}\rangle$ with $\langle\cdot\rangle$ denoting the average over the trial wave function.
This replacement is justified in the present calculations by the fact that the RDMC accounts for only a small part of the total energy.
After the replacement, $I^{\Delta\tau}_{\alpha\beta}(\bm R',\bm R)$ can be easily evaluated with the standard Trotter formula~\cite{Trotter1959Proc.Am.Math.Soc.545}.
Note that a similar recipe has been adopted in the AFDMC calculations to include the commutator terms of $3N$ interactions into the imaginary-time propagation ~\cite{Lonardoni2018Phys.Rev.Lett.122502,Lonardoni2018Phys.Rev.C044318}.

In the present RDMC calculations, a set of walkers is initially sampled from the RVMC wave function, and then a branching random walk algorithm is carried out~\cite{Carlson2015Rev.Mod.Phys.1067}.
The imaginary-time step is taken to be $\Delta\tau=5\times10^{-5}$ MeV$^{-1}$.
For each step, the diffusion and branching of the walkers are carried out according to the free and interaction terms of the short-time propagator in Eq.~(\ref{Eq.prop}), respectively.
The importance sampling is carried out as in the GFMC calculations~\cite{Pudliner1997Phys.Rev.C1720,Wiringa2000Phys.Rev.C014001}.
The mixed estimate of the energy is calculated every 20 propagation steps.

\begin{figure}[!htpb]
    \centering
	\includegraphics[width=0.45\textwidth]{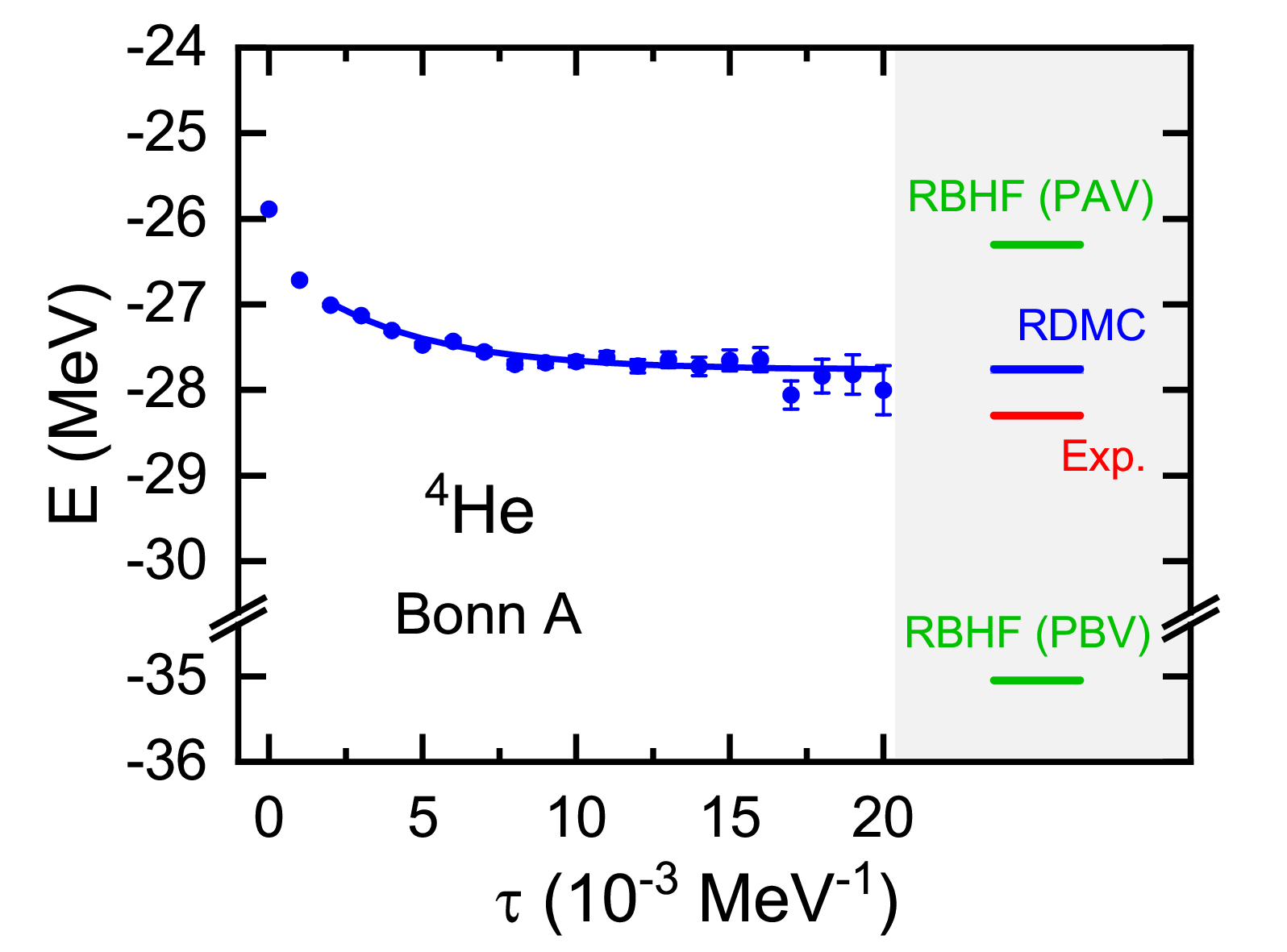}
	\caption{(Color online). The energy of $^4$He as a function of the imaginary time in the RDMC calculations with Bonn A.
The solid curve is an exponential fit to the Monte Carlo results.
Also shown are the RBHF results~\cite{Shen2017Phys.Rev.C014316} obtained with the projection before/after variation (PBV/PAV) approximations.
	}\label{fig2}
\end{figure}

Figure~\ref{fig2} depicts the energy of $^4$He as a function of the imaginary time in the RDMC calculation with Bonn A.
The energy drops rapidly within $\tau=2\times10^{-3}$ MeV$^{-1}$ by filtering out high excitations.
The ground-state energy is extracted from an exponential fit to the Monte Carlo results after $\tau=2\times10^{-3}$ MeV$^{-1}$, as shown by the solid curve.
The imaginary-time evolution lowers the RVMC energy by about 2 MeV.

We compare the present ground-state energy against the previous RBHF results with the same interaction Bonn A~\cite{Shen2017Phys.Rev.C014316}.
The RBHF method is a mean-field-based method, so the results are drastically influenced by the treatment of the center-of-mass motions, i.e., the projection before or after variation approximations.
In contrast, the present RDMC energy well reproduces the experimental value $E=-28.3$ MeV within $\sim2\%$.
This highlights the importance of the present virtually exact calculations.

\begin{figure}[!htpb]
    \centering
	\includegraphics[width=0.45\textwidth]{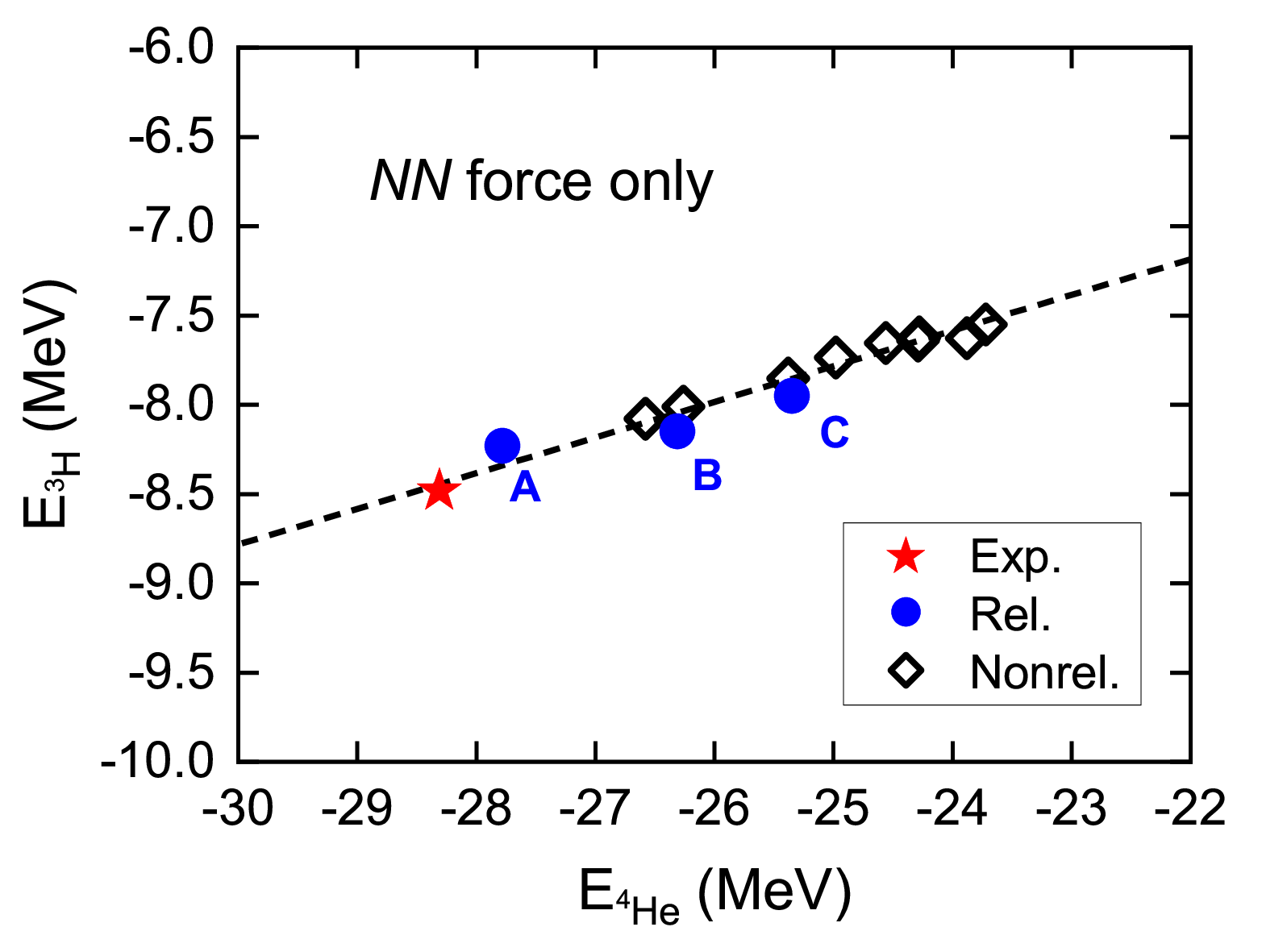}
	\caption{(Color online). The ground-state energies of $^3$H and $^4$He obtained from the present RDMC calculations with Bonn A, B, and C potentials, compared to the nonrelativistic results with various $NN$ forces (taken from Refs.~\cite{Nogga2000Phys.Rev.Lett.944947,Lynn2014Phys.Rev.Lett.192501,Binder2016Phys.Rev.C044002,Marcucci2020Front.Phys.69}).
	The dashed line is a linear fit to the nonrelativistic results (Tjon line).
}\label{fig3}
\end{figure}
{\it Relativistic \textit{ab initio} predictions.}---Figure~\ref{fig3} depicts the $^3$H and $^4$He ground-state energies obtained from the RDMC calculations with Bonn A, B, and C potentials.
The nonrelativistic results from virtually exact \textit{ab initio} methods~\cite{Nogga2000Phys.Rev.Lett.944947,Lynn2014Phys.Rev.Lett.192501,Binder2016Phys.Rev.C044002,Marcucci2020Front.Phys.69} with various $NN$ forces are also shown for comparison.
They cover nearly all existing classes of realistic $NN$ forces, including AV18~\cite{Wiringa1995Phys.Rev.C38}, CD-Bonn~\cite{Machleidt2001Phys.Rev.C024001}, Nijmegen I and II~\cite{Stoks1994Phys.Rev.C29502962}, and chiral $NN$ forces up to the third order ~\cite{Gezerlis2013Phys.Rev.Lett.032501},
the fourth order~\cite{Entem2003Phys.Rev.C041001,Piarulli2016Phys.Rev.C054007},
and the fifth order~\cite{Epelbaum2015Phys.Rev.Lett.122301,Entem2017Phys.Rev.C024004}.
For the chiral forces, the cutoff is either $\Lambda=500$ MeV in the momentum space or $R_0=1.0$ fm in the coordinate space.

The nonrelativistic results are systematically underbound for both $^3$H and $^4$He, so the $3N$ forces are often introduced and adjusted to the $^3$H or $^4$He ground-state energies.
In such an \textit{ad hoc} way, both the ground-state energies of $^3$H and $^4$He can be reproduced since they are correlated (Tjon line~\cite{Tjon1975Phys.Lett.B217220}).
In contrast to the nonrelativistic case, the relativistic results are much closer to the experimental values without any adjustment, especially for the results obtained with Bonn A.
In addition, the correlation between the $^3$H and $^4$He energies, namely the Tjon line, has also been revealed in the present relativistic calculations.

\begin{figure}[!htpb]
    \centering
	\includegraphics[width=0.45\textwidth]{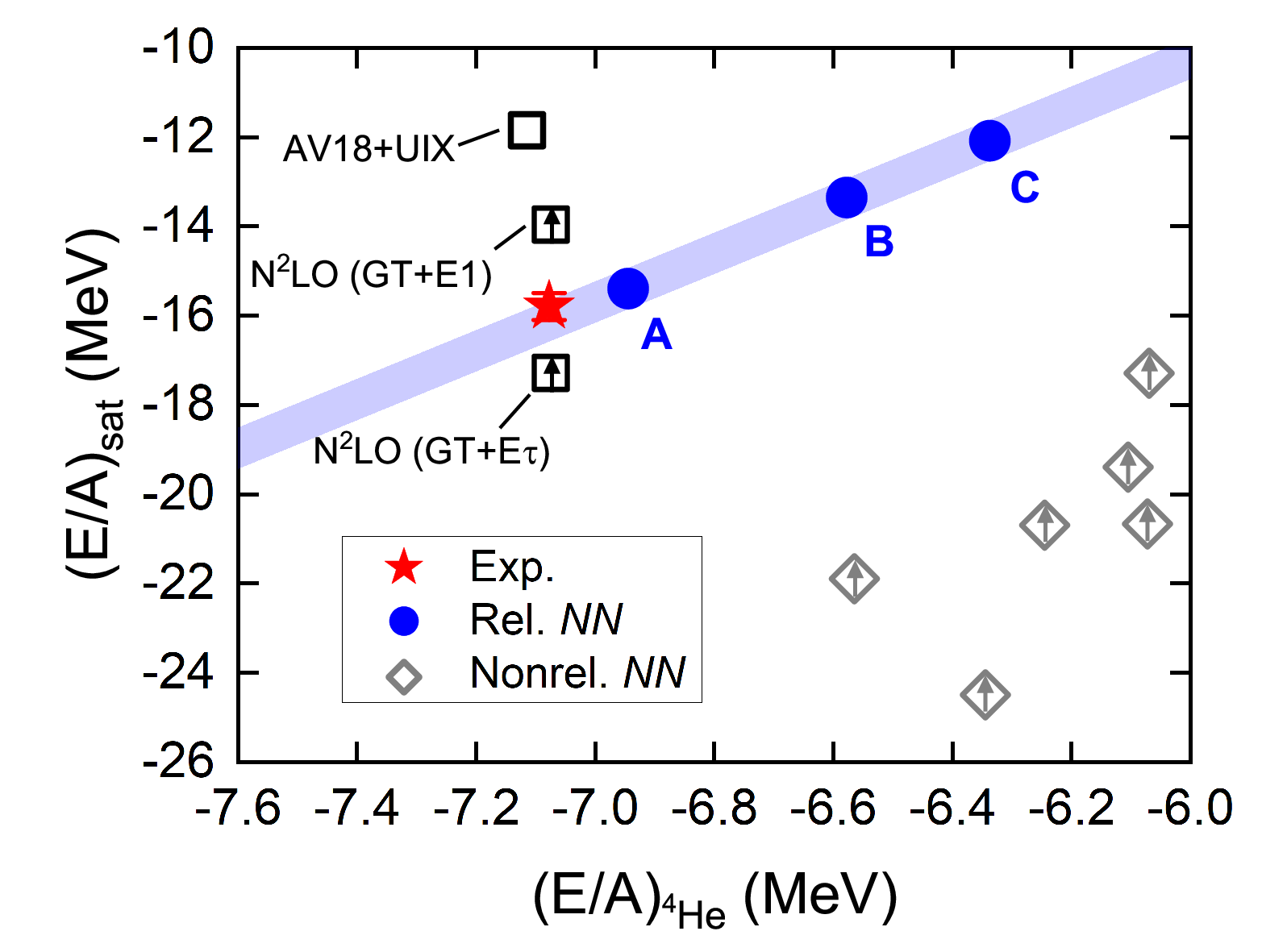}
	\caption{(Color online). Energy per nucleon for the $^4$He ground state and the saturation point of symmetric nuclear matter obtained from various \textit{ab initio} calculations.
The experimental value of $^4$He energy and the empirical value of the saturation energy~\cite{Margueron2018Phys.Rev.C025805} are shown for comparison.
The blue band is shown to guide the eyes.
The upper arrow indicates that the saturation density is overestimated by over $20\%$ compared to the empirical value~\cite{Margueron2018Phys.Rev.C025805}.
The nonrelativistic results for $^4$He and nuclear matter are respectively taken from Refs.~\cite{Nogga2000Phys.Rev.Lett.944947,Lynn2014Phys.Rev.Lett.192501,Lynn2016Phys.Rev.Lett.062501,Binder2016Phys.Rev.C044002,Marcucci2020Front.Phys.69}
and Refs.~\cite{Akmal1998Phys.Rev.C18041828,Li2006Phys.Rev.C047304,Hu2017Phys.Rev.C034307,Lonardoni2020Phys.Rev.Res.022033},
and the relativistic results for the nuclear matter are from the RBHF calculations~\cite{Wang2021Phys.Rev.C054319}.
	}\label{fig4}
\end{figure}

Figure~\ref{fig4} depicts various \textit{ab initio} results for the energy per nucleon of the $^4$He ground state and the saturation point of symmetric nuclear matter.
With only $NN$ forces, all the nonrelativistic calculations underbind the $^4$He ground state, but overbind the nuclear saturation energy and overestimate the corresponding saturation density.
The $^4$He energy can be reproduced by adjusting the $3N$ forces as in Refs.~\cite{Pudliner1995Phys.Rev.Lett.43964399,Lynn2016Phys.Rev.Lett.062501}.
However, with the same $3N$ forces, an accurate reproduction of the saturation properties is still a challenge.
As seen in Fig.~\ref{fig4}, the results obtained with AV18+UIX~\cite{Akmal1998Phys.Rev.C18041828} remarkably underbind the saturation energy, even though the saturation density is well reproduced by fit.
The results with the N$^2$LO chiral forces~\cite{Lonardoni2020Phys.Rev.Res.022033} somehow improve the saturation energy, depending on the choice of the $3N$ operator form ($E1$ or $E\tau$), while the corresponding saturation density always comes out too high.

On the contrary, the relativistic calculations with Bonn A provide the only result that can simultaneously reproduce the $^4$He energy and
the nuclear saturation energy and density.
One should keep in mind that the calculation is free of any adjustment since Bonn A is a high-precision two-nucleon potential.
For the results with Bonn B and C, the discrepancies against the experimental values are larger, mainly due to their larger tensor force components, which implies more repulsion in few- and many-body systems~\cite{Machleidt1989}.
However, in contrast to the nonrelativistic results, the relativistic results with Bonn A, B, and C exhibit regular correlations, i.e.,
the energy discrepancies for $^4$He and nuclear saturation are in the same direction.
This provides a quantitative understanding of the connection between the light nuclei and nuclear matter saturation properties.
It implies that one could reconcile light nuclei and nuclear matter in relativistic \textit{ab initio} calculations with only $NN$ force.

At this point, one may ask how large the $3N$-force effects should be in the relativistic \textit{ab initio} calculations.
A modern understanding of $3N$ forces is provided by chiral effective field theory, where the contributions of $NN$ forces are leading order, and the contributions of $3N$ forces are higher order corrections~\cite{Epelbaum2009Rev.Mod.Phys.17731825,Machleidt2011Phys.Rep.175,Hammer2020Rev.Mod.Phys.025004} and are expected to be small.
Note that the Bonn $NN$ force adopted in the present work is based on a phenomenological meson-exchange picture, which does not allow systematic improvements as well as the construction of consistent $3N$ forces.
Recently, significant progress has been made in constructing relativistic chiral forces in the relativistic frameworks~\cite{Epelbaum2012Phys.Lett.B338344,Ren2018Chin.Phys.C014103,Ren2022Phys.Rev.C034001,Lu2022Phys.Rev.Lett.142002}.
The relativistic QMC methods developed in this work pave the way for accurate \textit{ab initio} calculations of light and medium-mass nuclei with relativistic chiral $NN$ forces, as well as the construction of relativistic chiral $3N$ forces.

{\it Summary.}---We develop the relativistic variational Monte Carlo and relativistic diffusion Monte Carlo methods, and calculate the ground-state energies for $A\leq4$ nuclei using the Bonn force with an unprecedented high accuracy.
In contrast to the nonrelativistic results with $NN$ forces that underbind $A=3,4$ nuclei, the relativistic results are much closer to the experimental values without any adjustment, especially for the results obtained with Bonn A.
From the present results of $A\leq4$ nuclei and the previous results of nuclear matter with the same Bonn $NN$ forces~\cite{Wang2021Phys.Rev.C054319},
a correlation between the energies of light $A\leq4$ nuclei and nuclear saturation is revealed in the relativistic \textit{ab initio} calculations.
It highlights the importance of relativistic effects in a simultaneous description of light nuclei and nuclear matter.
It also provides a quantitative understanding of the connection between the light nuclei and nuclear matter saturation properties, which has been an outstanding problem in nuclear \textit{ab initio} calculations for decades.

\begin{acknowledgments}
We thank Evgeny Epelbaum for valuable discussions.
This work has been supported in part by the National Natural Science Foundation of China (Grants No. 123B2080, No. 12141501, No. 11935003), and the High-performance Computing Platform of Peking University.
We acknowledge the funding support from the State Key Laboratory of Nuclear Physics and Technology, Peking University (Grant No. NPT2023ZX03).
\end{acknowledgments}


\begin{thebibliography}{68}%
\makeatletter
\providecommand \@ifxundefined [1]{%
 \@ifx{#1\undefined}
}%
\providecommand \@ifnum [1]{%
 \ifnum #1\expandafter \@firstoftwo
 \else \expandafter \@secondoftwo
 \fi
}%
\providecommand \@ifx [1]{%
 \ifx #1\expandafter \@firstoftwo
 \else \expandafter \@secondoftwo
 \fi
}%
\providecommand \natexlab [1]{#1}%
\providecommand \enquote  [1]{``#1''}%
\providecommand \bibnamefont  [1]{#1}%
\providecommand \bibfnamefont [1]{#1}%
\providecommand \citenamefont [1]{#1}%
\providecommand \href@noop [0]{\@secondoftwo}%
\providecommand \href [0]{\begingroup \@sanitize@url \@href}%
\providecommand \@href[1]{\@@startlink{#1}\@@href}%
\providecommand \@@href[1]{\endgroup#1\@@endlink}%
\providecommand \@sanitize@url [0]{\catcode `\\12\catcode `\$12\catcode
  `\&12\catcode `\#12\catcode `\^12\catcode `\_12\catcode `\%12\relax}%
\providecommand \@@startlink[1]{}%
\providecommand \@@endlink[0]{}%
\providecommand \url  [0]{\begingroup\@sanitize@url \@url }%
\providecommand \@url [1]{\endgroup\@href {#1}{\urlprefix }}%
\providecommand \urlprefix  [0]{URL }%
\providecommand \Eprint [0]{\href }%
\providecommand \doibase [0]{https://doi.org/}%
\providecommand \selectlanguage [0]{\@gobble}%
\providecommand \bibinfo  [0]{\@secondoftwo}%
\providecommand \bibfield  [0]{\@secondoftwo}%
\providecommand \translation [1]{[#1]}%
\providecommand \BibitemOpen [0]{}%
\providecommand \bibitemStop [0]{}%
\providecommand \bibitemNoStop [0]{.\EOS\space}%
\providecommand \EOS [0]{\spacefactor3000\relax}%
\providecommand \BibitemShut  [1]{\csname bibitem#1\endcsname}%
\let\auto@bib@innerbib\@empty
\bibitem [{\citenamefont {Wiringa}\ \emph {et~al.}(1995)\citenamefont
  {Wiringa}, \citenamefont {Stoks},\ and\ \citenamefont
  {Schiavilla}}]{Wiringa1995Phys.Rev.C38}%
  \BibitemOpen
  \bibfield  {author} {\bibinfo {author} {\bibfnamefont {R.~B.}\ \bibnamefont
  {Wiringa}}, \bibinfo {author} {\bibfnamefont {V.~G.~J.}\ \bibnamefont
  {Stoks}},\ and\ \bibinfo {author} {\bibfnamefont {R.}~\bibnamefont
  {Schiavilla}},\ }\bibfield  {title} {\bibinfo {title} {{Accurate
  nucleon-nucleon potential with charge-independence breaking}},\ }\href
  {https://doi.org/10.1103/PhysRevC.51.38} {\bibfield  {journal} {\bibinfo
  {journal} {Phys. Rev. C}\ }\textbf {\bibinfo {volume} {51}},\ \bibinfo
  {pages} {38} (\bibinfo {year} {1995})}\BibitemShut {NoStop}%
\bibitem [{\citenamefont {Machleidt}(1989)}]{Machleidt1989}%
  \BibitemOpen
  \bibfield  {author} {\bibinfo {author} {\bibfnamefont {R.}~\bibnamefont
  {Machleidt}},\ }\bibinfo {title} {{The Meson Theory of Nuclear Forces and
  Nuclear Structure}},\ in\ \href {https://doi.org/10.1007/978-1-4613-9907-0_2}
  {\emph {\bibinfo {booktitle} {Advances in Nuclear Physics}}},\ \bibinfo
  {editor} {edited by\ \bibinfo {editor} {\bibfnamefont {J.~W.}\ \bibnamefont
  {Negele}}\ and\ \bibinfo {editor} {\bibfnamefont {E.}~\bibnamefont {Vogt}}}\
  (\bibinfo  {publisher} {Springer US},\ \bibinfo {address} {Boston, MA},\
  \bibinfo {year} {1989})\ pp.\ \bibinfo {pages} {189--376}\BibitemShut
  {NoStop}%
\bibitem [{\citenamefont {Machleidt}(2001)}]{Machleidt2001Phys.Rev.C024001}%
  \BibitemOpen
  \bibfield  {author} {\bibinfo {author} {\bibfnamefont {R.}~\bibnamefont
  {Machleidt}},\ }\bibfield  {title} {\bibinfo {title} {{High-precision,
  charge-dependent Bonn nucleon-nucleon potential}},\ }\href
  {https://doi.org/10.1103/PhysRevC.63.024001} {\bibfield  {journal} {\bibinfo
  {journal} {Phys. Rev. C}\ }\textbf {\bibinfo {volume} {63}},\ \bibinfo
  {pages} {024001} (\bibinfo {year} {2001})}\BibitemShut {NoStop}%
\bibitem [{\citenamefont {Pudliner}\ \emph {et~al.}(1995)\citenamefont
  {Pudliner}, \citenamefont {Pandharipande}, \citenamefont {Carlson},\ and\
  \citenamefont {Wiringa}}]{Pudliner1995Phys.Rev.Lett.43964399}%
  \BibitemOpen
  \bibfield  {author} {\bibinfo {author} {\bibfnamefont {B.~S.}\ \bibnamefont
  {Pudliner}}, \bibinfo {author} {\bibfnamefont {V.~R.}\ \bibnamefont
  {Pandharipande}}, \bibinfo {author} {\bibfnamefont {J.}~\bibnamefont
  {Carlson}},\ and\ \bibinfo {author} {\bibfnamefont {R.~B.}\ \bibnamefont
  {Wiringa}},\ }\bibfield  {title} {\bibinfo {title} {{Quantum Monte Carlo
  Calculations of $\mathit{A}\ensuremath{\le}6$ Nuclei}},\ }\href
  {https://doi.org/10.1103/PhysRevLett.74.4396} {\bibfield  {journal} {\bibinfo
   {journal} {Phys. Rev. Lett.}\ }\textbf {\bibinfo {volume} {74}},\ \bibinfo
  {pages} {4396} (\bibinfo {year} {1995})}\BibitemShut {NoStop}%
\bibitem [{\citenamefont {Navr\'atil}\ \emph {et~al.}(2000)\citenamefont
  {Navr\'atil}, \citenamefont {Vary},\ and\ \citenamefont
  {Barrett}}]{Navratil2000Phys.Rev.C054311}%
  \BibitemOpen
  \bibfield  {author} {\bibinfo {author} {\bibfnamefont {P.}~\bibnamefont
  {Navr\'atil}}, \bibinfo {author} {\bibfnamefont {J.~P.}\ \bibnamefont
  {Vary}},\ and\ \bibinfo {author} {\bibfnamefont {B.~R.}\ \bibnamefont
  {Barrett}},\ }\bibfield  {title} {\bibinfo {title} {{Large-basis ab initio
  no-core shell model and its application to ${}^{12}\mathbf{C}$}},\ }\href
  {https://doi.org/10.1103/PhysRevC.62.054311} {\bibfield  {journal} {\bibinfo
  {journal} {Phys. Rev. C}\ }\textbf {\bibinfo {volume} {62}},\ \bibinfo
  {pages} {054311} (\bibinfo {year} {2000})}\BibitemShut {NoStop}%
\bibitem [{\citenamefont {Epelbaum}\ \emph {et~al.}(2002)\citenamefont
  {Epelbaum}, \citenamefont {Nogga}, \citenamefont {Gl\"ockle}, \citenamefont
  {Kamada}, \citenamefont {Mei\ss{}ner},\ and\ \citenamefont
  {Wita\l{}a}}]{Epelbaum2002Phys.Rev.C064001}%
  \BibitemOpen
  \bibfield  {author} {\bibinfo {author} {\bibfnamefont {E.}~\bibnamefont
  {Epelbaum}}, \bibinfo {author} {\bibfnamefont {A.}~\bibnamefont {Nogga}},
  \bibinfo {author} {\bibfnamefont {W.}~\bibnamefont {Gl\"ockle}}, \bibinfo
  {author} {\bibfnamefont {H.}~\bibnamefont {Kamada}}, \bibinfo {author}
  {\bibfnamefont {U.-G.}\ \bibnamefont {Mei\ss{}ner}},\ and\ \bibinfo {author}
  {\bibfnamefont {H.}~\bibnamefont {Wita\l{}a}},\ }\bibfield  {title} {\bibinfo
  {title} {{Three-nucleon forces from chiral effective field theory}},\ }\href
  {https://doi.org/10.1103/PhysRevC.66.064001} {\bibfield  {journal} {\bibinfo
  {journal} {Phys. Rev. C}\ }\textbf {\bibinfo {volume} {66}},\ \bibinfo
  {pages} {064001} (\bibinfo {year} {2002})}\BibitemShut {NoStop}%
\bibitem [{\citenamefont {Navr\'atil}\ \emph {et~al.}(2007)\citenamefont
  {Navr\'atil}, \citenamefont {Gueorguiev}, \citenamefont {Vary}, \citenamefont
  {Ormand},\ and\ \citenamefont {Nogga}}]{Navratil2007Phys.Rev.Lett.042501}%
  \BibitemOpen
  \bibfield  {author} {\bibinfo {author} {\bibfnamefont {P.}~\bibnamefont
  {Navr\'atil}}, \bibinfo {author} {\bibfnamefont {V.~G.}\ \bibnamefont
  {Gueorguiev}}, \bibinfo {author} {\bibfnamefont {J.~P.}\ \bibnamefont
  {Vary}}, \bibinfo {author} {\bibfnamefont {W.~E.}\ \bibnamefont {Ormand}},\
  and\ \bibinfo {author} {\bibfnamefont {A.}~\bibnamefont {Nogga}},\ }\bibfield
   {title} {\bibinfo {title} {{Structure of $A=10$-$13$ Nuclei with Two- Plus
  Three-Nucleon Interactions from Chiral Effective Field Theory}},\ }\href
  {https://doi.org/10.1103/PhysRevLett.99.042501} {\bibfield  {journal}
  {\bibinfo  {journal} {Phys. Rev. Lett.}\ }\textbf {\bibinfo {volume} {99}},\
  \bibinfo {pages} {042501} (\bibinfo {year} {2007})}\BibitemShut {NoStop}%
\bibitem [{\citenamefont {Pieper}(2008)}]{Pieper2008AIPConf.Proc.143152}%
  \BibitemOpen
  \bibfield  {author} {\bibinfo {author} {\bibfnamefont {S.~C.}\ \bibnamefont
  {Pieper}},\ }\bibfield  {title} {\bibinfo {title} {{The Illinois Extension to
  the Fujita‐Miyazawa Three‐Nucleon Force}},\ }\href
  {https://doi.org/10.1063/1.2932280} {\bibfield  {journal} {\bibinfo
  {journal} {AIP Conf. Proc.}\ }\textbf {\bibinfo {volume} {1011}},\ \bibinfo
  {pages} {143} (\bibinfo {year} {2008})}\BibitemShut {NoStop}%
\bibitem [{\citenamefont {Gazit}\ \emph {et~al.}(2009)\citenamefont {Gazit},
  \citenamefont {Quaglioni},\ and\ \citenamefont
  {Navr\'atil}}]{Gazit2009Phys.Rev.Lett.102502}%
  \BibitemOpen
  \bibfield  {author} {\bibinfo {author} {\bibfnamefont {D.}~\bibnamefont
  {Gazit}}, \bibinfo {author} {\bibfnamefont {S.}~\bibnamefont {Quaglioni}},\
  and\ \bibinfo {author} {\bibfnamefont {P.}~\bibnamefont {Navr\'atil}},\
  }\bibfield  {title} {\bibinfo {title} {{Three-Nucleon Low-Energy Constants
  from the Consistency of Interactions and Currents in Chiral Effective Field
  Theory}},\ }\href {https://doi.org/10.1103/PhysRevLett.103.102502} {\bibfield
   {journal} {\bibinfo  {journal} {Phys. Rev. Lett.}\ }\textbf {\bibinfo
  {volume} {103}},\ \bibinfo {pages} {102502} (\bibinfo {year}
  {2009})}\BibitemShut {NoStop}%
\bibitem [{\citenamefont {Lynn}\ \emph {et~al.}(2016)\citenamefont {Lynn},
  \citenamefont {Tews}, \citenamefont {Carlson}, \citenamefont {Gandolfi},
  \citenamefont {Gezerlis}, \citenamefont {Schmidt},\ and\ \citenamefont
  {Schwenk}}]{Lynn2016Phys.Rev.Lett.062501}%
  \BibitemOpen
  \bibfield  {author} {\bibinfo {author} {\bibfnamefont {J.~E.}\ \bibnamefont
  {Lynn}}, \bibinfo {author} {\bibfnamefont {I.}~\bibnamefont {Tews}}, \bibinfo
  {author} {\bibfnamefont {J.}~\bibnamefont {Carlson}}, \bibinfo {author}
  {\bibfnamefont {S.}~\bibnamefont {Gandolfi}}, \bibinfo {author}
  {\bibfnamefont {A.}~\bibnamefont {Gezerlis}}, \bibinfo {author}
  {\bibfnamefont {K.~E.}\ \bibnamefont {Schmidt}},\ and\ \bibinfo {author}
  {\bibfnamefont {A.}~\bibnamefont {Schwenk}},\ }\bibfield  {title} {\bibinfo
  {title} {{Chiral Three-Nucleon Interactions in Light Nuclei,
  Neutron-$\ensuremath{\alpha}$ Scattering, and Neutron Matter}},\ }\href
  {https://doi.org/10.1103/PhysRevLett.116.062501} {\bibfield  {journal}
  {\bibinfo  {journal} {Phys. Rev. Lett.}\ }\textbf {\bibinfo {volume} {116}},\
  \bibinfo {pages} {062501} (\bibinfo {year} {2016})}\BibitemShut {NoStop}%
\bibitem [{\citenamefont {Piarulli}\ \emph {et~al.}(2018)\citenamefont
  {Piarulli}, \citenamefont {Baroni}, \citenamefont {Girlanda}, \citenamefont
  {Kievsky}, \citenamefont {Lovato}, \citenamefont {Lusk}, \citenamefont
  {Marcucci}, \citenamefont {Pieper}, \citenamefont {Schiavilla}, \citenamefont
  {Viviani},\ and\ \citenamefont {Wiringa}}]{Piarulli2018Phys.Rev.Lett.052503}%
  \BibitemOpen
  \bibfield  {author} {\bibinfo {author} {\bibfnamefont {M.}~\bibnamefont
  {Piarulli}}, \bibinfo {author} {\bibfnamefont {A.}~\bibnamefont {Baroni}},
  \bibinfo {author} {\bibfnamefont {L.}~\bibnamefont {Girlanda}}, \bibinfo
  {author} {\bibfnamefont {A.}~\bibnamefont {Kievsky}}, \bibinfo {author}
  {\bibfnamefont {A.}~\bibnamefont {Lovato}}, \bibinfo {author} {\bibfnamefont
  {E.}~\bibnamefont {Lusk}}, \bibinfo {author} {\bibfnamefont {L.~E.}\
  \bibnamefont {Marcucci}}, \bibinfo {author} {\bibfnamefont {S.~C.}\
  \bibnamefont {Pieper}}, \bibinfo {author} {\bibfnamefont {R.}~\bibnamefont
  {Schiavilla}}, \bibinfo {author} {\bibfnamefont {M.}~\bibnamefont
  {Viviani}},\ and\ \bibinfo {author} {\bibfnamefont {R.~B.}\ \bibnamefont
  {Wiringa}},\ }\bibfield  {title} {\bibinfo {title} {{Light-Nuclei Spectra
  from Chiral Dynamics}},\ }\href
  {https://doi.org/10.1103/PhysRevLett.120.052503} {\bibfield  {journal}
  {\bibinfo  {journal} {Phys. Rev. Lett.}\ }\textbf {\bibinfo {volume} {120}},\
  \bibinfo {pages} {052503} (\bibinfo {year} {2018})}\BibitemShut {NoStop}%
\bibitem [{\citenamefont {Epelbaum}\ \emph {et~al.}(2019)\citenamefont
  {Epelbaum}, \citenamefont {Golak}, \citenamefont {Hebeler}, \citenamefont
  {H\"uther}, \citenamefont {Kamada}, \citenamefont {Krebs}, \citenamefont
  {Maris}, \citenamefont {Mei\ss{}ner}, \citenamefont {Nogga}, \citenamefont
  {Roth}, \citenamefont {Skibi\ifmmode~\acute{n}\else \'{n}\fi{}ski},
  \citenamefont {Topolnicki}, \citenamefont {Vary}, \citenamefont {Vobig},\
  and\ \citenamefont {Wita\l{}a}}]{Epelbaum2019Phys.Rev.C024313}%
  \BibitemOpen
  \bibfield  {author} {\bibinfo {author} {\bibfnamefont {E.}~\bibnamefont
  {Epelbaum}}, \bibinfo {author} {\bibfnamefont {J.}~\bibnamefont {Golak}},
  \bibinfo {author} {\bibfnamefont {K.}~\bibnamefont {Hebeler}}, \bibinfo
  {author} {\bibfnamefont {T.}~\bibnamefont {H\"uther}}, \bibinfo {author}
  {\bibfnamefont {H.}~\bibnamefont {Kamada}}, \bibinfo {author} {\bibfnamefont
  {H.}~\bibnamefont {Krebs}}, \bibinfo {author} {\bibfnamefont
  {P.}~\bibnamefont {Maris}}, \bibinfo {author} {\bibfnamefont {U.-G.}\
  \bibnamefont {Mei\ss{}ner}}, \bibinfo {author} {\bibfnamefont
  {A.}~\bibnamefont {Nogga}}, \bibinfo {author} {\bibfnamefont
  {R.}~\bibnamefont {Roth}}, \bibinfo {author} {\bibfnamefont {R.}~\bibnamefont
  {Skibi\ifmmode~\acute{n}\else \'{n}\fi{}ski}}, \bibinfo {author}
  {\bibfnamefont {K.}~\bibnamefont {Topolnicki}}, \bibinfo {author}
  {\bibfnamefont {J.~P.}\ \bibnamefont {Vary}}, \bibinfo {author}
  {\bibfnamefont {K.}~\bibnamefont {Vobig}},\ and\ \bibinfo {author}
  {\bibfnamefont {H.}~\bibnamefont {Wita\l{}a}} (\bibinfo {collaboration}
  {LENPIC Collaboration}),\ }\bibfield  {title} {\bibinfo {title} {{Few- and
  many-nucleon systems with semilocal coordinate-space regularized chiral two-
  and three-body forces}},\ }\href {https://doi.org/10.1103/PhysRevC.99.024313}
  {\bibfield  {journal} {\bibinfo  {journal} {Phys. Rev. C}\ }\textbf {\bibinfo
  {volume} {99}},\ \bibinfo {pages} {024313} (\bibinfo {year}
  {2019})}\BibitemShut {NoStop}%
\bibitem [{\citenamefont {Coraggio}\ \emph {et~al.}(2014)\citenamefont
  {Coraggio}, \citenamefont {Holt}, \citenamefont {Itaco}, \citenamefont
  {Machleidt}, \citenamefont {Marcucci},\ and\ \citenamefont
  {Sammarruca}}]{Coraggio2014Phys.Rev.C044321}%
  \BibitemOpen
  \bibfield  {author} {\bibinfo {author} {\bibfnamefont {L.}~\bibnamefont
  {Coraggio}}, \bibinfo {author} {\bibfnamefont {J.~W.}\ \bibnamefont {Holt}},
  \bibinfo {author} {\bibfnamefont {N.}~\bibnamefont {Itaco}}, \bibinfo
  {author} {\bibfnamefont {R.}~\bibnamefont {Machleidt}}, \bibinfo {author}
  {\bibfnamefont {L.~E.}\ \bibnamefont {Marcucci}},\ and\ \bibinfo {author}
  {\bibfnamefont {F.}~\bibnamefont {Sammarruca}},\ }\bibfield  {title}
  {\bibinfo {title} {{Nuclear-matter equation of state with consistent two- and
  three-body perturbative chiral interactions}},\ }\href
  {https://doi.org/10.1103/PhysRevC.89.044321} {\bibfield  {journal} {\bibinfo
  {journal} {Phys. Rev. C}\ }\textbf {\bibinfo {volume} {89}},\ \bibinfo
  {pages} {044321} (\bibinfo {year} {2014})}\BibitemShut {NoStop}%
\bibitem [{\citenamefont {Sammarruca}\ \emph {et~al.}(2015)\citenamefont
  {Sammarruca}, \citenamefont {Coraggio}, \citenamefont {Holt}, \citenamefont
  {Itaco}, \citenamefont {Machleidt},\ and\ \citenamefont
  {Marcucci}}]{Sammarruca2015Phys.Rev.C054311}%
  \BibitemOpen
  \bibfield  {author} {\bibinfo {author} {\bibfnamefont {F.}~\bibnamefont
  {Sammarruca}}, \bibinfo {author} {\bibfnamefont {L.}~\bibnamefont
  {Coraggio}}, \bibinfo {author} {\bibfnamefont {J.~W.}\ \bibnamefont {Holt}},
  \bibinfo {author} {\bibfnamefont {N.}~\bibnamefont {Itaco}}, \bibinfo
  {author} {\bibfnamefont {R.}~\bibnamefont {Machleidt}},\ and\ \bibinfo
  {author} {\bibfnamefont {L.~E.}\ \bibnamefont {Marcucci}},\ }\bibfield
  {title} {\bibinfo {title} {{Toward order-by-order calculations of the nuclear
  and neutron matter equations of state in chiral effective field theory}},\
  }\href {https://doi.org/10.1103/PhysRevC.91.054311} {\bibfield  {journal}
  {\bibinfo  {journal} {Phys. Rev. C}\ }\textbf {\bibinfo {volume} {91}},\
  \bibinfo {pages} {054311} (\bibinfo {year} {2015})}\BibitemShut {NoStop}%
\bibitem [{\citenamefont {Lonardoni}\ \emph {et~al.}(2020)\citenamefont
  {Lonardoni}, \citenamefont {Tews}, \citenamefont {Gandolfi},\ and\
  \citenamefont {Carlson}}]{Lonardoni2020Phys.Rev.Res.022033}%
  \BibitemOpen
  \bibfield  {author} {\bibinfo {author} {\bibfnamefont {D.}~\bibnamefont
  {Lonardoni}}, \bibinfo {author} {\bibfnamefont {I.}~\bibnamefont {Tews}},
  \bibinfo {author} {\bibfnamefont {S.}~\bibnamefont {Gandolfi}},\ and\
  \bibinfo {author} {\bibfnamefont {J.}~\bibnamefont {Carlson}},\ }\bibfield
  {title} {\bibinfo {title} {{Nuclear and neutron-star matter from local chiral
  interactions}},\ }\href {https://doi.org/10.1103/PhysRevResearch.2.022033}
  {\bibfield  {journal} {\bibinfo  {journal} {Phys. Rev. Res.}\ }\textbf
  {\bibinfo {volume} {2}},\ \bibinfo {pages} {022033} (\bibinfo {year}
  {2020})}\BibitemShut {NoStop}%
\bibitem [{\citenamefont {Hebeler}(2021)}]{Hebeler2021Phys.Rep.1116}%
  \BibitemOpen
  \bibfield  {author} {\bibinfo {author} {\bibfnamefont {K.}~\bibnamefont
  {Hebeler}},\ }\bibfield  {title} {\bibinfo {title} {{Three-nucleon forces:
  Implementation and applications to atomic nuclei and dense matter}},\ }\href
  {https://doi.org/https://doi.org/10.1016/j.physrep.2020.08.009} {\bibfield
  {journal} {\bibinfo  {journal} {Phys. Rep.}\ }\textbf {\bibinfo {volume}
  {890}},\ \bibinfo {pages} {1} (\bibinfo {year} {2021})},\ \bibinfo {note}
  {three-nucleon forces: Implementation and applications to atomic nuclei and
  dense matter}\BibitemShut {NoStop}%
\bibitem [{\citenamefont {Ekstr\"om}\ \emph {et~al.}(2015)\citenamefont
  {Ekstr\"om}, \citenamefont {Jansen}, \citenamefont {Wendt}, \citenamefont
  {Hagen}, \citenamefont {Papenbrock}, \citenamefont {Carlsson}, \citenamefont
  {Forss\'en}, \citenamefont {Hjorth-Jensen}, \citenamefont {Navr\'atil},\ and\
  \citenamefont {Nazarewicz}}]{Ekstroem2015Phys.Rev.C051301}%
  \BibitemOpen
  \bibfield  {author} {\bibinfo {author} {\bibfnamefont {A.}~\bibnamefont
  {Ekstr\"om}}, \bibinfo {author} {\bibfnamefont {G.~R.}\ \bibnamefont
  {Jansen}}, \bibinfo {author} {\bibfnamefont {K.~A.}\ \bibnamefont {Wendt}},
  \bibinfo {author} {\bibfnamefont {G.}~\bibnamefont {Hagen}}, \bibinfo
  {author} {\bibfnamefont {T.}~\bibnamefont {Papenbrock}}, \bibinfo {author}
  {\bibfnamefont {B.~D.}\ \bibnamefont {Carlsson}}, \bibinfo {author}
  {\bibfnamefont {C.}~\bibnamefont {Forss\'en}}, \bibinfo {author}
  {\bibfnamefont {M.}~\bibnamefont {Hjorth-Jensen}}, \bibinfo {author}
  {\bibfnamefont {P.}~\bibnamefont {Navr\'atil}},\ and\ \bibinfo {author}
  {\bibfnamefont {W.}~\bibnamefont {Nazarewicz}},\ }\bibfield  {title}
  {\bibinfo {title} {{Accurate nuclear radii and binding energies from a chiral
  interaction}},\ }\href {https://doi.org/10.1103/PhysRevC.91.051301}
  {\bibfield  {journal} {\bibinfo  {journal} {Phys. Rev. C}\ }\textbf {\bibinfo
  {volume} {91}},\ \bibinfo {pages} {051301} (\bibinfo {year}
  {2015})}\BibitemShut {NoStop}%
\bibitem [{\citenamefont {Drischler}\ \emph {et~al.}(2019)\citenamefont
  {Drischler}, \citenamefont {Hebeler},\ and\ \citenamefont
  {Schwenk}}]{Drischler2019Phys.Rev.Lett.042501}%
  \BibitemOpen
  \bibfield  {author} {\bibinfo {author} {\bibfnamefont {C.}~\bibnamefont
  {Drischler}}, \bibinfo {author} {\bibfnamefont {K.}~\bibnamefont {Hebeler}},\
  and\ \bibinfo {author} {\bibfnamefont {A.}~\bibnamefont {Schwenk}},\
  }\bibfield  {title} {\bibinfo {title} {{Chiral Interactions up to
  Next-to-Next-to-Next-to-Leading Order and Nuclear Saturation}},\ }\href
  {https://doi.org/10.1103/PhysRevLett.122.042501} {\bibfield  {journal}
  {\bibinfo  {journal} {Phys. Rev. Lett.}\ }\textbf {\bibinfo {volume} {122}},\
  \bibinfo {pages} {042501} (\bibinfo {year} {2019})}\BibitemShut {NoStop}%
\bibitem [{\citenamefont {Anastasio}\ \emph {et~al.}(1980)\citenamefont
  {Anastasio}, \citenamefont {Celenza},\ and\ \citenamefont
  {Shakin}}]{Anastasio1980Phys.Rev.Lett.20962099}%
  \BibitemOpen
  \bibfield  {author} {\bibinfo {author} {\bibfnamefont {M.~R.}\ \bibnamefont
  {Anastasio}}, \bibinfo {author} {\bibfnamefont {L.~S.}\ \bibnamefont
  {Celenza}},\ and\ \bibinfo {author} {\bibfnamefont {C.~M.}\ \bibnamefont
  {Shakin}},\ }\bibfield  {title} {\bibinfo {title} {{Nuclear Saturation as a
  Relativistic Effect}},\ }\href {https://doi.org/10.1103/PhysRevLett.45.2096}
  {\bibfield  {journal} {\bibinfo  {journal} {Phys. Rev. Lett.}\ }\textbf
  {\bibinfo {volume} {45}},\ \bibinfo {pages} {2096} (\bibinfo {year}
  {1980})}\BibitemShut {NoStop}%
\bibitem [{\citenamefont {Brockmann}\ and\ \citenamefont
  {Machleidt}(1990)}]{Brockmann1990Phys.Rev.C19651980}%
  \BibitemOpen
  \bibfield  {author} {\bibinfo {author} {\bibfnamefont {R.}~\bibnamefont
  {Brockmann}}\ and\ \bibinfo {author} {\bibfnamefont {R.}~\bibnamefont
  {Machleidt}},\ }\bibfield  {title} {\bibinfo {title} {{Relativistic nuclear
  structure. I. Nuclear matter}},\ }\href
  {https://doi.org/10.1103/PhysRevC.42.1965} {\bibfield  {journal} {\bibinfo
  {journal} {Phys. Rev. C}\ }\textbf {\bibinfo {volume} {42}},\ \bibinfo
  {pages} {1965} (\bibinfo {year} {1990})}\BibitemShut {NoStop}%
\bibitem [{\citenamefont {Sammarruca}\ \emph {et~al.}(2012)\citenamefont
  {Sammarruca}, \citenamefont {Chen}, \citenamefont {Coraggio}, \citenamefont
  {Itaco},\ and\ \citenamefont {Machleidt}}]{Sammarruca2012Phys.Rev.C054317}%
  \BibitemOpen
  \bibfield  {author} {\bibinfo {author} {\bibfnamefont {F.}~\bibnamefont
  {Sammarruca}}, \bibinfo {author} {\bibfnamefont {B.}~\bibnamefont {Chen}},
  \bibinfo {author} {\bibfnamefont {L.}~\bibnamefont {Coraggio}}, \bibinfo
  {author} {\bibfnamefont {N.}~\bibnamefont {Itaco}},\ and\ \bibinfo {author}
  {\bibfnamefont {R.}~\bibnamefont {Machleidt}},\ }\bibfield  {title} {\bibinfo
  {title} {{Dirac-Brueckner-Hartree-Fock versus chiral effective field
  theory}},\ }\href {https://doi.org/10.1103/PhysRevC.86.054317} {\bibfield
  {journal} {\bibinfo  {journal} {Phys. Rev. C}\ }\textbf {\bibinfo {volume}
  {86}},\ \bibinfo {pages} {054317} (\bibinfo {year} {2012})}\BibitemShut
  {NoStop}%
\bibitem [{\citenamefont {Wang}\ \emph {et~al.}(2021)\citenamefont {Wang},
  \citenamefont {Zhao}, \citenamefont {Ring},\ and\ \citenamefont
  {Meng}}]{Wang2021Phys.Rev.C054319}%
  \BibitemOpen
  \bibfield  {author} {\bibinfo {author} {\bibfnamefont {S.}~\bibnamefont
  {Wang}}, \bibinfo {author} {\bibfnamefont {Q.}~\bibnamefont {Zhao}}, \bibinfo
  {author} {\bibfnamefont {P.}~\bibnamefont {Ring}},\ and\ \bibinfo {author}
  {\bibfnamefont {J.}~\bibnamefont {Meng}},\ }\bibfield  {title} {\bibinfo
  {title} {{Nuclear matter in relativistic Brueckner-Hartree-Fock theory with
  Bonn potential in the full Dirac space}},\ }\href
  {https://doi.org/10.1103/physrevc.103.054319} {\bibfield  {journal} {\bibinfo
   {journal} {Phys. Rev. C}\ }\textbf {\bibinfo {volume} {103}},\ \bibinfo
  {pages} {054319} (\bibinfo {year} {2021})}\BibitemShut {NoStop}%
\bibitem [{\citenamefont {Shen}\ \emph {et~al.}(2016)\citenamefont {Shen},
  \citenamefont {Hu}, \citenamefont {Liang}, \citenamefont {Meng},
  \citenamefont {Ring},\ and\ \citenamefont
  {Zhang}}]{Shen2016Chin.Phys.Lett.102103}%
  \BibitemOpen
  \bibfield  {author} {\bibinfo {author} {\bibfnamefont {S.-H.}\ \bibnamefont
  {Shen}}, \bibinfo {author} {\bibfnamefont {J.-N.}\ \bibnamefont {Hu}},
  \bibinfo {author} {\bibfnamefont {H.-Z.}\ \bibnamefont {Liang}}, \bibinfo
  {author} {\bibfnamefont {J.}~\bibnamefont {Meng}}, \bibinfo {author}
  {\bibfnamefont {P.}~\bibnamefont {Ring}},\ and\ \bibinfo {author}
  {\bibfnamefont {S.-Q.}\ \bibnamefont {Zhang}},\ }\bibfield  {title} {\bibinfo
  {title} {{Relativistic Brueckner{\textemdash}Hartree{\textemdash}Fock Theory
  for Finite Nuclei}},\ }\href {https://doi.org/10.1088/0256-307x/33/10/102103}
  {\bibfield  {journal} {\bibinfo  {journal} {Chin. Phys. Lett.}\ }\textbf
  {\bibinfo {volume} {33}},\ \bibinfo {pages} {102103} (\bibinfo {year}
  {2016})}\BibitemShut {NoStop}%
\bibitem [{\citenamefont {Shen}\ \emph {et~al.}(2017)\citenamefont {Shen},
  \citenamefont {Liang}, \citenamefont {Meng}, \citenamefont {Ring},\ and\
  \citenamefont {Zhang}}]{Shen2017Phys.Rev.C014316}%
  \BibitemOpen
  \bibfield  {author} {\bibinfo {author} {\bibfnamefont {S.}~\bibnamefont
  {Shen}}, \bibinfo {author} {\bibfnamefont {H.}~\bibnamefont {Liang}},
  \bibinfo {author} {\bibfnamefont {J.}~\bibnamefont {Meng}}, \bibinfo {author}
  {\bibfnamefont {P.}~\bibnamefont {Ring}},\ and\ \bibinfo {author}
  {\bibfnamefont {S.}~\bibnamefont {Zhang}},\ }\bibfield  {title} {\bibinfo
  {title} {{Fully self-consistent relativistic Brueckner-Hartree-Fock theory
  for finite nuclei}},\ }\href {https://doi.org/10.1103/PhysRevC.96.014316}
  {\bibfield  {journal} {\bibinfo  {journal} {Phys. Rev. C}\ }\textbf {\bibinfo
  {volume} {96}},\ \bibinfo {pages} {014316} (\bibinfo {year}
  {2017})}\BibitemShut {NoStop}%
\bibitem [{\citenamefont {Shen}\ \emph {et~al.}(2019)\citenamefont {Shen},
  \citenamefont {Liang}, \citenamefont {Long}, \citenamefont {Meng},\ and\
  \citenamefont {Ring}}]{Shen2019Prog.Part.Nucl.Phys.103713}%
  \BibitemOpen
  \bibfield  {author} {\bibinfo {author} {\bibfnamefont {S.}~\bibnamefont
  {Shen}}, \bibinfo {author} {\bibfnamefont {H.}~\bibnamefont {Liang}},
  \bibinfo {author} {\bibfnamefont {W.~H.}\ \bibnamefont {Long}}, \bibinfo
  {author} {\bibfnamefont {J.}~\bibnamefont {Meng}},\ and\ \bibinfo {author}
  {\bibfnamefont {P.}~\bibnamefont {Ring}},\ }\bibfield  {title} {\bibinfo
  {title} {{Towards an ab initio covariant density functional theory for
  nuclear structure}},\ }\href
  {https://doi.org/https://doi.org/10.1016/j.ppnp.2019.103713} {\bibfield
  {journal} {\bibinfo  {journal} {Prog. Part. Nucl. Phys.}\ }\textbf {\bibinfo
  {volume} {109}},\ \bibinfo {pages} {103713} (\bibinfo {year}
  {2019})}\BibitemShut {NoStop}%
\bibitem [{\citenamefont {Stadler}\ and\ \citenamefont
  {Gross}(1997)}]{Stadler1997Phys.Rev.Lett.2629}%
  \BibitemOpen
  \bibfield  {author} {\bibinfo {author} {\bibfnamefont {A.}~\bibnamefont
  {Stadler}}\ and\ \bibinfo {author} {\bibfnamefont {F.}~\bibnamefont
  {Gross}},\ }\bibfield  {title} {\bibinfo {title} {{Relativistic Calculation
  of the Triton Binding Energy and Its Implications}},\ }\href
  {https://doi.org/10.1103/PhysRevLett.78.26} {\bibfield  {journal} {\bibinfo
  {journal} {Phys. Rev. Lett.}\ }\textbf {\bibinfo {volume} {78}},\ \bibinfo
  {pages} {26} (\bibinfo {year} {1997})}\BibitemShut {NoStop}%
\bibitem [{\citenamefont {Sammarruca}\ \emph {et~al.}(1992)\citenamefont
  {Sammarruca}, \citenamefont {Xu},\ and\ \citenamefont
  {Machleidt}}]{Sammarruca1992Phys.Rev.C16361641}%
  \BibitemOpen
  \bibfield  {author} {\bibinfo {author} {\bibfnamefont {F.}~\bibnamefont
  {Sammarruca}}, \bibinfo {author} {\bibfnamefont {D.~P.}\ \bibnamefont {Xu}},\
  and\ \bibinfo {author} {\bibfnamefont {R.}~\bibnamefont {Machleidt}},\
  }\bibfield  {title} {\bibinfo {title} {{Relativistic corrections to the
  triton binding energy}},\ }\href {https://doi.org/10.1103/PhysRevC.46.1636}
  {\bibfield  {journal} {\bibinfo  {journal} {Phys. Rev. C}\ }\textbf {\bibinfo
  {volume} {46}},\ \bibinfo {pages} {1636} (\bibinfo {year}
  {1992})}\BibitemShut {NoStop}%
\bibitem [{\citenamefont {Carlson}\ \emph {et~al.}(2015)\citenamefont
  {Carlson}, \citenamefont {Gandolfi}, \citenamefont {Pederiva}, \citenamefont
  {Pieper}, \citenamefont {Schiavilla}, \citenamefont {Schmidt},\ and\
  \citenamefont {Wiringa}}]{Carlson2015Rev.Mod.Phys.1067}%
  \BibitemOpen
  \bibfield  {author} {\bibinfo {author} {\bibfnamefont {J.}~\bibnamefont
  {Carlson}}, \bibinfo {author} {\bibfnamefont {S.}~\bibnamefont {Gandolfi}},
  \bibinfo {author} {\bibfnamefont {F.}~\bibnamefont {Pederiva}}, \bibinfo
  {author} {\bibfnamefont {S.~C.}\ \bibnamefont {Pieper}}, \bibinfo {author}
  {\bibfnamefont {R.}~\bibnamefont {Schiavilla}}, \bibinfo {author}
  {\bibfnamefont {K.~E.}\ \bibnamefont {Schmidt}},\ and\ \bibinfo {author}
  {\bibfnamefont {R.~B.}\ \bibnamefont {Wiringa}},\ }\bibfield  {title}
  {\bibinfo {title} {{Quantum Monte Carlo methods for nuclear physics}},\
  }\href {https://doi.org/10.1103/RevModPhys.87.1067} {\bibfield  {journal}
  {\bibinfo  {journal} {Rev. Mod. Phys.}\ }\textbf {\bibinfo {volume} {87}},\
  \bibinfo {pages} {1067} (\bibinfo {year} {2015})}\BibitemShut {NoStop}%
\bibitem [{\citenamefont {Lynn}\ \emph {et~al.}(2019)\citenamefont {Lynn},
  \citenamefont {Tews}, \citenamefont {Gandolfi},\ and\ \citenamefont
  {Lovato}}]{Lynn2019Ann.Rev.Nucl.Part.Sci.279305}%
  \BibitemOpen
  \bibfield  {author} {\bibinfo {author} {\bibfnamefont {J.}~\bibnamefont
  {Lynn}}, \bibinfo {author} {\bibfnamefont {I.}~\bibnamefont {Tews}}, \bibinfo
  {author} {\bibfnamefont {S.}~\bibnamefont {Gandolfi}},\ and\ \bibinfo
  {author} {\bibfnamefont {A.}~\bibnamefont {Lovato}},\ }\bibfield  {title}
  {\bibinfo {title} {{Quantum Monte Carlo Methods in Nuclear Physics: Recent
  Advances}},\ }\href {https://doi.org/10.1146/annurev-nucl-101918-023600}
  {\bibfield  {journal} {\bibinfo  {journal} {Ann. Rev. Nucl. Part. Sci.}\
  }\textbf {\bibinfo {volume} {69}},\ \bibinfo {pages} {279} (\bibinfo {year}
  {2019})},\ \Eprint
  {https://arxiv.org/abs/https://doi.org/10.1146/annurev-nucl-101918-023600}
  {https://doi.org/10.1146/annurev-nucl-101918-023600} \BibitemShut {NoStop}%
\bibitem [{\citenamefont {Gandolfi}\ \emph {et~al.}(2020)\citenamefont
  {Gandolfi}, \citenamefont {Lonardoni}, \citenamefont {Lovato},\ and\
  \citenamefont {Piarulli}}]{Gandolfi2020Front.Phys.}%
  \BibitemOpen
  \bibfield  {author} {\bibinfo {author} {\bibfnamefont {S.}~\bibnamefont
  {Gandolfi}}, \bibinfo {author} {\bibfnamefont {D.}~\bibnamefont {Lonardoni}},
  \bibinfo {author} {\bibfnamefont {A.}~\bibnamefont {Lovato}},\ and\ \bibinfo
  {author} {\bibfnamefont {M.}~\bibnamefont {Piarulli}},\ }\bibfield  {title}
  {\bibinfo {title} {{Atomic Nuclei From Quantum Monte Carlo Calculations With
  Chiral EFT Interactions}},\ }\bibfield  {journal} {\bibinfo  {journal}
  {Front. Phys.}\ }\textbf {\bibinfo {volume} {8}},\ \href
  {https://doi.org/10.3389/fphy.2020.00117} {10.3389/fphy.2020.00117} (\bibinfo
  {year} {2020})\BibitemShut {NoStop}%
\bibitem [{\citenamefont {Wiringa}\ and\ \citenamefont
  {Pieper}(2002)}]{Wiringa2002Phys.Rev.Lett.182501}%
  \BibitemOpen
  \bibfield  {author} {\bibinfo {author} {\bibfnamefont {R.~B.}\ \bibnamefont
  {Wiringa}}\ and\ \bibinfo {author} {\bibfnamefont {S.~C.}\ \bibnamefont
  {Pieper}},\ }\bibfield  {title} {\bibinfo {title} {{Evolution of Nuclear
  Spectra with Nuclear Forces}},\ }\href
  {https://doi.org/10.1103/PhysRevLett.89.182501} {\bibfield  {journal}
  {\bibinfo  {journal} {Phys. Rev. Lett.}\ }\textbf {\bibinfo {volume} {89}},\
  \bibinfo {pages} {182501} (\bibinfo {year} {2002})}\BibitemShut {NoStop}%
\bibitem [{\citenamefont {Lovato}\ \emph {et~al.}(2013)\citenamefont {Lovato},
  \citenamefont {Gandolfi}, \citenamefont {Butler}, \citenamefont {Carlson},
  \citenamefont {Lusk}, \citenamefont {Pieper},\ and\ \citenamefont
  {Schiavilla}}]{Lovato2013Phys.Rev.Lett.092501}%
  \BibitemOpen
  \bibfield  {author} {\bibinfo {author} {\bibfnamefont {A.}~\bibnamefont
  {Lovato}}, \bibinfo {author} {\bibfnamefont {S.}~\bibnamefont {Gandolfi}},
  \bibinfo {author} {\bibfnamefont {R.}~\bibnamefont {Butler}}, \bibinfo
  {author} {\bibfnamefont {J.}~\bibnamefont {Carlson}}, \bibinfo {author}
  {\bibfnamefont {E.}~\bibnamefont {Lusk}}, \bibinfo {author} {\bibfnamefont
  {S.~C.}\ \bibnamefont {Pieper}},\ and\ \bibinfo {author} {\bibfnamefont
  {R.}~\bibnamefont {Schiavilla}},\ }\bibfield  {title} {\bibinfo {title}
  {{Charge Form Factor and Sum Rules of Electromagnetic Response Functions in
  $^{12}\mathbf{C}$}},\ }\href {https://doi.org/10.1103/PhysRevLett.111.092501}
  {\bibfield  {journal} {\bibinfo  {journal} {Phys. Rev. Lett.}\ }\textbf
  {\bibinfo {volume} {111}},\ \bibinfo {pages} {092501} (\bibinfo {year}
  {2013})}\BibitemShut {NoStop}%
\bibitem [{\citenamefont {Gandolfi}\ \emph
  {et~al.}(2007{\natexlab{a}})\citenamefont {Gandolfi}, \citenamefont
  {Pederiva}, \citenamefont {Fantoni},\ and\ \citenamefont
  {Schmidt}}]{Gandolfi2007Phys.Rev.Lett.022507}%
  \BibitemOpen
  \bibfield  {author} {\bibinfo {author} {\bibfnamefont {S.}~\bibnamefont
  {Gandolfi}}, \bibinfo {author} {\bibfnamefont {F.}~\bibnamefont {Pederiva}},
  \bibinfo {author} {\bibfnamefont {S.}~\bibnamefont {Fantoni}},\ and\ \bibinfo
  {author} {\bibfnamefont {K.~E.}\ \bibnamefont {Schmidt}},\ }\bibfield
  {title} {\bibinfo {title} {{Auxiliary Field Diffusion Monte Carlo Calculation
  of Nuclei with $A\ensuremath{\le}40$ with Tensor Interactions}},\ }\href
  {https://doi.org/10.1103/PhysRevLett.99.022507} {\bibfield  {journal}
  {\bibinfo  {journal} {Phys. Rev. Lett.}\ }\textbf {\bibinfo {volume} {99}},\
  \bibinfo {pages} {022507} (\bibinfo {year} {2007}{\natexlab{a}})}\BibitemShut
  {NoStop}%
\bibitem [{\citenamefont {Gandolfi}\ \emph
  {et~al.}(2007{\natexlab{b}})\citenamefont {Gandolfi}, \citenamefont
  {Pederiva}, \citenamefont {Fantoni},\ and\ \citenamefont
  {Schmidt}}]{Gandolfi2007Phys.Rev.Lett.102503}%
  \BibitemOpen
  \bibfield  {author} {\bibinfo {author} {\bibfnamefont {S.}~\bibnamefont
  {Gandolfi}}, \bibinfo {author} {\bibfnamefont {F.}~\bibnamefont {Pederiva}},
  \bibinfo {author} {\bibfnamefont {S.}~\bibnamefont {Fantoni}},\ and\ \bibinfo
  {author} {\bibfnamefont {K.~E.}\ \bibnamefont {Schmidt}},\ }\bibfield
  {title} {\bibinfo {title} {{Quantum Monte Carlo Calculations of Symmetric
  Nuclear Matter}},\ }\href {https://doi.org/10.1103/PhysRevLett.98.102503}
  {\bibfield  {journal} {\bibinfo  {journal} {Phys. Rev. Lett.}\ }\textbf
  {\bibinfo {volume} {98}},\ \bibinfo {pages} {102503} (\bibinfo {year}
  {2007}{\natexlab{b}})}\BibitemShut {NoStop}%
\bibitem [{\citenamefont {Lonardoni}\ \emph
  {et~al.}(2018{\natexlab{a}})\citenamefont {Lonardoni}, \citenamefont
  {Carlson}, \citenamefont {Gandolfi}, \citenamefont {Lynn}, \citenamefont
  {Schmidt}, \citenamefont {Schwenk},\ and\ \citenamefont
  {Wang}}]{Lonardoni2018Phys.Rev.Lett.122502}%
  \BibitemOpen
  \bibfield  {author} {\bibinfo {author} {\bibfnamefont {D.}~\bibnamefont
  {Lonardoni}}, \bibinfo {author} {\bibfnamefont {J.}~\bibnamefont {Carlson}},
  \bibinfo {author} {\bibfnamefont {S.}~\bibnamefont {Gandolfi}}, \bibinfo
  {author} {\bibfnamefont {J.~E.}\ \bibnamefont {Lynn}}, \bibinfo {author}
  {\bibfnamefont {K.~E.}\ \bibnamefont {Schmidt}}, \bibinfo {author}
  {\bibfnamefont {A.}~\bibnamefont {Schwenk}},\ and\ \bibinfo {author}
  {\bibfnamefont {X.~B.}\ \bibnamefont {Wang}},\ }\bibfield  {title} {\bibinfo
  {title} {{Properties of Nuclei up to $A=16$ using Local Chiral
  Interactions}},\ }\href {https://doi.org/10.1103/PhysRevLett.120.122502}
  {\bibfield  {journal} {\bibinfo  {journal} {Phys. Rev. Lett.}\ }\textbf
  {\bibinfo {volume} {120}},\ \bibinfo {pages} {122502} (\bibinfo {year}
  {2018}{\natexlab{a}})}\BibitemShut {NoStop}%
\bibitem [{\citenamefont {Carlson}\ \emph {et~al.}(1993)\citenamefont
  {Carlson}, \citenamefont {Pandharipande},\ and\ \citenamefont
  {Schiavilla}}]{Carlson1993Phys.Rev.C484}%
  \BibitemOpen
  \bibfield  {author} {\bibinfo {author} {\bibfnamefont {J.}~\bibnamefont
  {Carlson}}, \bibinfo {author} {\bibfnamefont {V.~R.}\ \bibnamefont
  {Pandharipande}},\ and\ \bibinfo {author} {\bibfnamefont {R.}~\bibnamefont
  {Schiavilla}},\ }\bibfield  {title} {\bibinfo {title} {{Variational Monte
  Carlo calculations of $^{3}\mathrm{H}$ and $^{4}\mathrm{He}$ with a
  relativistic Hamiltonian}},\ }\href {https://doi.org/10.1103/PhysRevC.47.484}
  {\bibfield  {journal} {\bibinfo  {journal} {Phys. Rev. C}\ }\textbf {\bibinfo
  {volume} {47}},\ \bibinfo {pages} {484} (\bibinfo {year} {1993})}\BibitemShut
  {NoStop}%
\bibitem [{\citenamefont {Metropolis}\ \emph {et~al.}(1953)\citenamefont
  {Metropolis}, \citenamefont {Rosenbluth}, \citenamefont {Rosenbluth},
  \citenamefont {Teller},\ and\ \citenamefont
  {Teller}}]{Metropolis1953TheJournalofChemicalPhysics10871092}%
  \BibitemOpen
  \bibfield  {author} {\bibinfo {author} {\bibfnamefont {N.}~\bibnamefont
  {Metropolis}}, \bibinfo {author} {\bibfnamefont {A.~W.}\ \bibnamefont
  {Rosenbluth}}, \bibinfo {author} {\bibfnamefont {M.~N.}\ \bibnamefont
  {Rosenbluth}}, \bibinfo {author} {\bibfnamefont {A.~H.}\ \bibnamefont
  {Teller}},\ and\ \bibinfo {author} {\bibfnamefont {E.}~\bibnamefont
  {Teller}},\ }\bibfield  {title} {\bibinfo {title} {{Equation of State
  Calculations by Fast Computing Machines}},\ }\href
  {https://doi.org/10.1063/1.1699114} {\bibfield  {journal} {\bibinfo
  {journal} {The Journal of Chemical Physics}\ }\textbf {\bibinfo {volume}
  {21}},\ \bibinfo {pages} {1087} (\bibinfo {year} {1953})},\ \Eprint
  {https://arxiv.org/abs/https://doi.org/10.1063/1.1699114}
  {https://doi.org/10.1063/1.1699114} \BibitemShut {NoStop}%
\bibitem [{\citenamefont {Fahy}\ \emph {et~al.}(1988)\citenamefont {Fahy},
  \citenamefont {Wang},\ and\ \citenamefont
  {Louie}}]{Fahy1988Phys.Rev.Lett.1631}%
  \BibitemOpen
  \bibfield  {author} {\bibinfo {author} {\bibfnamefont {S.}~\bibnamefont
  {Fahy}}, \bibinfo {author} {\bibfnamefont {X.~W.}\ \bibnamefont {Wang}},\
  and\ \bibinfo {author} {\bibfnamefont {S.~G.}\ \bibnamefont {Louie}},\
  }\bibfield  {title} {\bibinfo {title} {Variational quantum monte carlo
  nonlocal pseudopotential approach to solids: Cohesive and structural
  properties of diamond},\ }\href {https://doi.org/10.1103/PhysRevLett.61.1631}
  {\bibfield  {journal} {\bibinfo  {journal} {Phys. Rev. Lett.}\ }\textbf
  {\bibinfo {volume} {61}},\ \bibinfo {pages} {1631} (\bibinfo {year}
  {1988})}\BibitemShut {NoStop}%
\bibitem [{\citenamefont {Fahy}\ \emph {et~al.}(1990)\citenamefont {Fahy},
  \citenamefont {Wang},\ and\ \citenamefont {Louie}}]{Fahy1990Phys.Rev.B3503}%
  \BibitemOpen
  \bibfield  {author} {\bibinfo {author} {\bibfnamefont {S.}~\bibnamefont
  {Fahy}}, \bibinfo {author} {\bibfnamefont {X.~W.}\ \bibnamefont {Wang}},\
  and\ \bibinfo {author} {\bibfnamefont {S.~G.}\ \bibnamefont {Louie}},\
  }\bibfield  {title} {\bibinfo {title} {{Variational quantum Monte Carlo
  nonlocal pseudopotential approach to solids: Formulation and application to
  diamond, graphite, and silicon}},\ }\href
  {https://doi.org/10.1103/PhysRevB.42.3503} {\bibfield  {journal} {\bibinfo
  {journal} {Phys. Rev. B}\ }\textbf {\bibinfo {volume} {42}},\ \bibinfo
  {pages} {3503} (\bibinfo {year} {1990})}\BibitemShut {NoStop}%
\bibitem [{\citenamefont {Adams}\ \emph {et~al.}(2021)\citenamefont {Adams},
  \citenamefont {Carleo}, \citenamefont {Lovato},\ and\ \citenamefont
  {Rocco}}]{Adams2021Phys.Rev.Lett.022502}%
  \BibitemOpen
  \bibfield  {author} {\bibinfo {author} {\bibfnamefont {C.}~\bibnamefont
  {Adams}}, \bibinfo {author} {\bibfnamefont {G.}~\bibnamefont {Carleo}},
  \bibinfo {author} {\bibfnamefont {A.}~\bibnamefont {Lovato}},\ and\ \bibinfo
  {author} {\bibfnamefont {N.}~\bibnamefont {Rocco}},\ }\bibfield  {title}
  {\bibinfo {title} {{Variational Monte Carlo Calculations of
  $A\ensuremath{\le}4$ Nuclei with an Artificial Neural-Network Correlator
  Ansatz}},\ }\href {https://doi.org/10.1103/PhysRevLett.127.022502} {\bibfield
   {journal} {\bibinfo  {journal} {Phys. Rev. Lett.}\ }\textbf {\bibinfo
  {volume} {127}},\ \bibinfo {pages} {022502} (\bibinfo {year}
  {2021})}\BibitemShut {NoStop}%
\bibitem [{\citenamefont {Yang}\ and\ \citenamefont
  {Zhao}(2022)}]{Yang2022Phys.Lett.B137587}%
  \BibitemOpen
  \bibfield  {author} {\bibinfo {author} {\bibfnamefont {Y.}~\bibnamefont
  {Yang}}\ and\ \bibinfo {author} {\bibfnamefont {P.}~\bibnamefont {Zhao}},\
  }\bibfield  {title} {\bibinfo {title} {{A consistent description of the
  relativistic effects and three-body interactions in atomic nuclei}},\ }\href
  {https://doi.org/https://doi.org/10.1016/j.physletb.2022.137587} {\bibfield
  {journal} {\bibinfo  {journal} {Phys. Lett. B}\ }\textbf {\bibinfo {volume}
  {835}},\ \bibinfo {pages} {137587} (\bibinfo {year} {2022})}\BibitemShut
  {NoStop}%
\bibitem [{\citenamefont {Yang}\ and\ \citenamefont
  {Zhao}(2023)}]{Yang2023Phys.Rev.C034320}%
  \BibitemOpen
  \bibfield  {author} {\bibinfo {author} {\bibfnamefont {Y.~L.}\ \bibnamefont
  {Yang}}\ and\ \bibinfo {author} {\bibfnamefont {P.~W.}\ \bibnamefont
  {Zhao}},\ }\bibfield  {title} {\bibinfo {title} {{Deep-neural-network
  approach to solving the ab initio nuclear structure problem}},\ }\href
  {https://doi.org/10.1103/PhysRevC.107.034320} {\bibfield  {journal} {\bibinfo
   {journal} {Phys. Rev. C}\ }\textbf {\bibinfo {volume} {107}},\ \bibinfo
  {pages} {034320} (\bibinfo {year} {2023})}\BibitemShut {NoStop}%
\bibitem [{\citenamefont {Trotter}(1959)}]{Trotter1959Proc.Am.Math.Soc.545}%
  \BibitemOpen
  \bibfield  {author} {\bibinfo {author} {\bibfnamefont {H.~F.}\ \bibnamefont
  {Trotter}},\ }\bibfield  {title} {\bibinfo {title} {{ON THE PRODUCT OF
  SEMI-GROUPS OF OPERATORS}},\ }\href
  {https://doi.org/http://dx.doi.org/10.1090/S0002-9939-1959-0108732-6}
  {\bibfield  {journal} {\bibinfo  {journal} {Proc. Am. Math. Soc.}\ }\textbf
  {\bibinfo {volume} {10}},\ \bibinfo {pages} {545} (\bibinfo {year}
  {1959})}\BibitemShut {NoStop}%
\bibitem [{\citenamefont {Lonardoni}\ \emph
  {et~al.}(2018{\natexlab{b}})\citenamefont {Lonardoni}, \citenamefont
  {Gandolfi}, \citenamefont {Lynn}, \citenamefont {Petrie}, \citenamefont
  {Carlson}, \citenamefont {Schmidt},\ and\ \citenamefont
  {Schwenk}}]{Lonardoni2018Phys.Rev.C044318}%
  \BibitemOpen
  \bibfield  {author} {\bibinfo {author} {\bibfnamefont {D.}~\bibnamefont
  {Lonardoni}}, \bibinfo {author} {\bibfnamefont {S.}~\bibnamefont {Gandolfi}},
  \bibinfo {author} {\bibfnamefont {J.~E.}\ \bibnamefont {Lynn}}, \bibinfo
  {author} {\bibfnamefont {C.}~\bibnamefont {Petrie}}, \bibinfo {author}
  {\bibfnamefont {J.}~\bibnamefont {Carlson}}, \bibinfo {author} {\bibfnamefont
  {K.~E.}\ \bibnamefont {Schmidt}},\ and\ \bibinfo {author} {\bibfnamefont
  {A.}~\bibnamefont {Schwenk}},\ }\bibfield  {title} {\bibinfo {title}
  {{Auxiliary field diffusion Monte Carlo calculations of light and medium-mass
  nuclei with local chiral interactions}},\ }\href
  {https://doi.org/10.1103/PhysRevC.97.044318} {\bibfield  {journal} {\bibinfo
  {journal} {Phys. Rev. C}\ }\textbf {\bibinfo {volume} {97}},\ \bibinfo
  {pages} {044318} (\bibinfo {year} {2018}{\natexlab{b}})}\BibitemShut
  {NoStop}%
\bibitem [{\citenamefont {Pudliner}\ \emph {et~al.}(1997)\citenamefont
  {Pudliner}, \citenamefont {Pandharipande}, \citenamefont {Carlson},
  \citenamefont {Pieper},\ and\ \citenamefont
  {Wiringa}}]{Pudliner1997Phys.Rev.C1720}%
  \BibitemOpen
  \bibfield  {author} {\bibinfo {author} {\bibfnamefont {B.~S.}\ \bibnamefont
  {Pudliner}}, \bibinfo {author} {\bibfnamefont {V.~R.}\ \bibnamefont
  {Pandharipande}}, \bibinfo {author} {\bibfnamefont {J.}~\bibnamefont
  {Carlson}}, \bibinfo {author} {\bibfnamefont {S.~C.}\ \bibnamefont
  {Pieper}},\ and\ \bibinfo {author} {\bibfnamefont {R.~B.}\ \bibnamefont
  {Wiringa}},\ }\bibfield  {title} {\bibinfo {title} {{Quantum Monte Carlo
  calculations of nuclei with $A<7$}},\ }\href
  {https://doi.org/10.1103/PhysRevC.56.1720} {\bibfield  {journal} {\bibinfo
  {journal} {Phys. Rev. C}\ }\textbf {\bibinfo {volume} {56}},\ \bibinfo
  {pages} {1720} (\bibinfo {year} {1997})}\BibitemShut {NoStop}%
\bibitem [{\citenamefont {Wiringa}\ \emph {et~al.}(2000)\citenamefont
  {Wiringa}, \citenamefont {Pieper}, \citenamefont {Carlson},\ and\
  \citenamefont {Pandharipande}}]{Wiringa2000Phys.Rev.C014001}%
  \BibitemOpen
  \bibfield  {author} {\bibinfo {author} {\bibfnamefont {R.~B.}\ \bibnamefont
  {Wiringa}}, \bibinfo {author} {\bibfnamefont {S.~C.}\ \bibnamefont {Pieper}},
  \bibinfo {author} {\bibfnamefont {J.}~\bibnamefont {Carlson}},\ and\ \bibinfo
  {author} {\bibfnamefont {V.~R.}\ \bibnamefont {Pandharipande}},\ }\bibfield
  {title} {\bibinfo {title} {{Quantum Monte Carlo calculations of $A=8$
  nuclei}},\ }\href {https://doi.org/10.1103/PhysRevC.62.014001} {\bibfield
  {journal} {\bibinfo  {journal} {Phys. Rev. C}\ }\textbf {\bibinfo {volume}
  {62}},\ \bibinfo {pages} {014001} (\bibinfo {year} {2000})}\BibitemShut
  {NoStop}%
\bibitem [{\citenamefont {Nogga}\ \emph {et~al.}(2000)\citenamefont {Nogga},
  \citenamefont {Kamada},\ and\ \citenamefont
  {Gl\"ockle}}]{Nogga2000Phys.Rev.Lett.944947}%
  \BibitemOpen
  \bibfield  {author} {\bibinfo {author} {\bibfnamefont {A.}~\bibnamefont
  {Nogga}}, \bibinfo {author} {\bibfnamefont {H.}~\bibnamefont {Kamada}},\ and\
  \bibinfo {author} {\bibfnamefont {W.}~\bibnamefont {Gl\"ockle}},\ }\bibfield
  {title} {\bibinfo {title} {{Modern Nuclear Force Predictions for the
  $\mathit{\ensuremath{\alpha}}$ Particle}},\ }\href
  {https://doi.org/10.1103/PhysRevLett.85.944} {\bibfield  {journal} {\bibinfo
  {journal} {Phys. Rev. Lett.}\ }\textbf {\bibinfo {volume} {85}},\ \bibinfo
  {pages} {944} (\bibinfo {year} {2000})}\BibitemShut {NoStop}%
\bibitem [{\citenamefont {Lynn}\ \emph {et~al.}(2014)\citenamefont {Lynn},
  \citenamefont {Carlson}, \citenamefont {Epelbaum}, \citenamefont {Gandolfi},
  \citenamefont {Gezerlis},\ and\ \citenamefont
  {Schwenk}}]{Lynn2014Phys.Rev.Lett.192501}%
  \BibitemOpen
  \bibfield  {author} {\bibinfo {author} {\bibfnamefont {J.~E.}\ \bibnamefont
  {Lynn}}, \bibinfo {author} {\bibfnamefont {J.}~\bibnamefont {Carlson}},
  \bibinfo {author} {\bibfnamefont {E.}~\bibnamefont {Epelbaum}}, \bibinfo
  {author} {\bibfnamefont {S.}~\bibnamefont {Gandolfi}}, \bibinfo {author}
  {\bibfnamefont {A.}~\bibnamefont {Gezerlis}},\ and\ \bibinfo {author}
  {\bibfnamefont {A.}~\bibnamefont {Schwenk}},\ }\bibfield  {title} {\bibinfo
  {title} {{Quantum Monte Carlo Calculations of Light Nuclei Using Chiral
  Potentials}},\ }\href {https://doi.org/10.1103/PhysRevLett.113.192501}
  {\bibfield  {journal} {\bibinfo  {journal} {Phys. Rev. Lett.}\ }\textbf
  {\bibinfo {volume} {113}},\ \bibinfo {pages} {192501} (\bibinfo {year}
  {2014})}\BibitemShut {NoStop}%
\bibitem [{\citenamefont {Binder}\ \emph {et~al.}(2016)\citenamefont {Binder},
  \citenamefont {Calci}, \citenamefont {Epelbaum}, \citenamefont {Furnstahl},
  \citenamefont {Golak}, \citenamefont {Hebeler}, \citenamefont {Kamada},
  \citenamefont {Krebs}, \citenamefont {Langhammer}, \citenamefont {Liebig},
  \citenamefont {Maris}, \citenamefont {Mei\ss{}ner}, \citenamefont {Minossi},
  \citenamefont {Nogga}, \citenamefont {Potter}, \citenamefont {Roth},
  \citenamefont {Skibi\ifmmode~\acute{n}\else \'{n}\fi{}ski}, \citenamefont
  {Topolnicki}, \citenamefont {Vary},\ and\ \citenamefont
  {Wita\l{}a}}]{Binder2016Phys.Rev.C044002}%
  \BibitemOpen
  \bibfield  {author} {\bibinfo {author} {\bibfnamefont {S.}~\bibnamefont
  {Binder}}, \bibinfo {author} {\bibfnamefont {A.}~\bibnamefont {Calci}},
  \bibinfo {author} {\bibfnamefont {E.}~\bibnamefont {Epelbaum}}, \bibinfo
  {author} {\bibfnamefont {R.~J.}\ \bibnamefont {Furnstahl}}, \bibinfo {author}
  {\bibfnamefont {J.}~\bibnamefont {Golak}}, \bibinfo {author} {\bibfnamefont
  {K.}~\bibnamefont {Hebeler}}, \bibinfo {author} {\bibfnamefont
  {H.}~\bibnamefont {Kamada}}, \bibinfo {author} {\bibfnamefont
  {H.}~\bibnamefont {Krebs}}, \bibinfo {author} {\bibfnamefont
  {J.}~\bibnamefont {Langhammer}}, \bibinfo {author} {\bibfnamefont
  {S.}~\bibnamefont {Liebig}}, \bibinfo {author} {\bibfnamefont
  {P.}~\bibnamefont {Maris}}, \bibinfo {author} {\bibfnamefont {U.-G.}\
  \bibnamefont {Mei\ss{}ner}}, \bibinfo {author} {\bibfnamefont
  {D.}~\bibnamefont {Minossi}}, \bibinfo {author} {\bibfnamefont
  {A.}~\bibnamefont {Nogga}}, \bibinfo {author} {\bibfnamefont
  {H.}~\bibnamefont {Potter}}, \bibinfo {author} {\bibfnamefont
  {R.}~\bibnamefont {Roth}}, \bibinfo {author} {\bibfnamefont {R.}~\bibnamefont
  {Skibi\ifmmode~\acute{n}\else \'{n}\fi{}ski}}, \bibinfo {author}
  {\bibfnamefont {K.}~\bibnamefont {Topolnicki}}, \bibinfo {author}
  {\bibfnamefont {J.~P.}\ \bibnamefont {Vary}},\ and\ \bibinfo {author}
  {\bibfnamefont {H.}~\bibnamefont {Wita\l{}a}} (\bibinfo {collaboration}
  {LENPIC Collaboration}),\ }\bibfield  {title} {\bibinfo {title} {{Few-nucleon
  systems with state-of-the-art chiral nucleon-nucleon forces}},\ }\href
  {https://doi.org/10.1103/PhysRevC.93.044002} {\bibfield  {journal} {\bibinfo
  {journal} {Phys. Rev. C}\ }\textbf {\bibinfo {volume} {93}},\ \bibinfo
  {pages} {044002} (\bibinfo {year} {2016})}\BibitemShut {NoStop}%
\bibitem [{\citenamefont {Marcucci}\ \emph {et~al.}(2020)\citenamefont
  {Marcucci}, \citenamefont {Dohet-Eraly}, \citenamefont {Girlanda},
  \citenamefont {Gnech}, \citenamefont {Kievsky},\ and\ \citenamefont
  {Viviani}}]{Marcucci2020Front.Phys.69}%
  \BibitemOpen
  \bibfield  {author} {\bibinfo {author} {\bibfnamefont {L.~E.}\ \bibnamefont
  {Marcucci}}, \bibinfo {author} {\bibfnamefont {J.}~\bibnamefont
  {Dohet-Eraly}}, \bibinfo {author} {\bibfnamefont {L.}~\bibnamefont
  {Girlanda}}, \bibinfo {author} {\bibfnamefont {A.}~\bibnamefont {Gnech}},
  \bibinfo {author} {\bibfnamefont {A.}~\bibnamefont {Kievsky}},\ and\ \bibinfo
  {author} {\bibfnamefont {M.}~\bibnamefont {Viviani}},\ }\bibfield  {title}
  {\bibinfo {title} {{The Hyperspherical Harmonics Method: A Tool for Testing
  and Improving Nuclear Interaction Models}},\ }\href
  {https://doi.org/10.3389/fphy.2020.00069} {\bibfield  {journal} {\bibinfo
  {journal} {Front. Phys.}\ }\textbf {\bibinfo {volume} {8}},\ \bibinfo {pages}
  {69} (\bibinfo {year} {2020})}\BibitemShut {NoStop}%
\bibitem [{\citenamefont {Stoks}\ \emph {et~al.}(1994)\citenamefont {Stoks},
  \citenamefont {Klomp}, \citenamefont {Terheggen},\ and\ \citenamefont
  {de~Swart}}]{Stoks1994Phys.Rev.C29502962}%
  \BibitemOpen
  \bibfield  {author} {\bibinfo {author} {\bibfnamefont {V.~G.~J.}\
  \bibnamefont {Stoks}}, \bibinfo {author} {\bibfnamefont {R.~A.~M.}\
  \bibnamefont {Klomp}}, \bibinfo {author} {\bibfnamefont {C.~P.~F.}\
  \bibnamefont {Terheggen}},\ and\ \bibinfo {author} {\bibfnamefont {J.~J.}\
  \bibnamefont {de~Swart}},\ }\bibfield  {title} {\bibinfo {title}
  {{Construction of high-quality NN potential models}},\ }\href
  {https://doi.org/10.1103/PhysRevC.49.2950} {\bibfield  {journal} {\bibinfo
  {journal} {Phys. Rev. C}\ }\textbf {\bibinfo {volume} {49}},\ \bibinfo
  {pages} {2950} (\bibinfo {year} {1994})}\BibitemShut {NoStop}%
\bibitem [{\citenamefont {Gezerlis}\ \emph {et~al.}(2013)\citenamefont
  {Gezerlis}, \citenamefont {Tews}, \citenamefont {Epelbaum}, \citenamefont
  {Gandolfi}, \citenamefont {Hebeler}, \citenamefont {Nogga},\ and\
  \citenamefont {Schwenk}}]{Gezerlis2013Phys.Rev.Lett.032501}%
  \BibitemOpen
  \bibfield  {author} {\bibinfo {author} {\bibfnamefont {A.}~\bibnamefont
  {Gezerlis}}, \bibinfo {author} {\bibfnamefont {I.}~\bibnamefont {Tews}},
  \bibinfo {author} {\bibfnamefont {E.}~\bibnamefont {Epelbaum}}, \bibinfo
  {author} {\bibfnamefont {S.}~\bibnamefont {Gandolfi}}, \bibinfo {author}
  {\bibfnamefont {K.}~\bibnamefont {Hebeler}}, \bibinfo {author} {\bibfnamefont
  {A.}~\bibnamefont {Nogga}},\ and\ \bibinfo {author} {\bibfnamefont
  {A.}~\bibnamefont {Schwenk}},\ }\bibfield  {title} {\bibinfo {title}
  {{Quantum Monte Carlo Calculations with Chiral Effective Field Theory
  Interactions}},\ }\href {https://doi.org/10.1103/PhysRevLett.111.032501}
  {\bibfield  {journal} {\bibinfo  {journal} {Phys. Rev. Lett.}\ }\textbf
  {\bibinfo {volume} {111}},\ \bibinfo {pages} {032501} (\bibinfo {year}
  {2013})}\BibitemShut {NoStop}%
\bibitem [{\citenamefont {Entem}\ and\ \citenamefont
  {Machleidt}(2003)}]{Entem2003Phys.Rev.C041001}%
  \BibitemOpen
  \bibfield  {author} {\bibinfo {author} {\bibfnamefont {D.~R.}\ \bibnamefont
  {Entem}}\ and\ \bibinfo {author} {\bibfnamefont {R.}~\bibnamefont
  {Machleidt}},\ }\bibfield  {title} {\bibinfo {title} {{Accurate
  charge-dependent nucleon-nucleon potential at fourth order of chiral
  perturbation theory}},\ }\href {https://doi.org/10.1103/PhysRevC.68.041001}
  {\bibfield  {journal} {\bibinfo  {journal} {Phys. Rev. C}\ }\textbf {\bibinfo
  {volume} {68}},\ \bibinfo {pages} {041001} (\bibinfo {year}
  {2003})}\BibitemShut {NoStop}%
\bibitem [{\citenamefont {Piarulli}\ \emph {et~al.}(2016)\citenamefont
  {Piarulli}, \citenamefont {Girlanda}, \citenamefont {Schiavilla},
  \citenamefont {Kievsky}, \citenamefont {Lovato}, \citenamefont {Marcucci},
  \citenamefont {Pieper}, \citenamefont {Viviani},\ and\ \citenamefont
  {Wiringa}}]{Piarulli2016Phys.Rev.C054007}%
  \BibitemOpen
  \bibfield  {author} {\bibinfo {author} {\bibfnamefont {M.}~\bibnamefont
  {Piarulli}}, \bibinfo {author} {\bibfnamefont {L.}~\bibnamefont {Girlanda}},
  \bibinfo {author} {\bibfnamefont {R.}~\bibnamefont {Schiavilla}}, \bibinfo
  {author} {\bibfnamefont {A.}~\bibnamefont {Kievsky}}, \bibinfo {author}
  {\bibfnamefont {A.}~\bibnamefont {Lovato}}, \bibinfo {author} {\bibfnamefont
  {L.~E.}\ \bibnamefont {Marcucci}}, \bibinfo {author} {\bibfnamefont {S.~C.}\
  \bibnamefont {Pieper}}, \bibinfo {author} {\bibfnamefont {M.}~\bibnamefont
  {Viviani}},\ and\ \bibinfo {author} {\bibfnamefont {R.~B.}\ \bibnamefont
  {Wiringa}},\ }\bibfield  {title} {\bibinfo {title} {{Local chiral potentials
  with $\mathrm{\ensuremath{\Delta}}$-intermediate states and the structure of
  light nuclei}},\ }\href {https://doi.org/10.1103/PhysRevC.94.054007}
  {\bibfield  {journal} {\bibinfo  {journal} {Phys. Rev. C}\ }\textbf {\bibinfo
  {volume} {94}},\ \bibinfo {pages} {054007} (\bibinfo {year}
  {2016})}\BibitemShut {NoStop}%
\bibitem [{\citenamefont {Epelbaum}\ \emph {et~al.}(2015)\citenamefont
  {Epelbaum}, \citenamefont {Krebs},\ and\ \citenamefont
  {Mei\ss{}ner}}]{Epelbaum2015Phys.Rev.Lett.122301}%
  \BibitemOpen
  \bibfield  {author} {\bibinfo {author} {\bibfnamefont {E.}~\bibnamefont
  {Epelbaum}}, \bibinfo {author} {\bibfnamefont {H.}~\bibnamefont {Krebs}},\
  and\ \bibinfo {author} {\bibfnamefont {U.-G.}\ \bibnamefont {Mei\ss{}ner}},\
  }\bibfield  {title} {\bibinfo {title} {{Precision Nucleon-Nucleon Potential
  at Fifth Order in the Chiral Expansion}},\ }\href
  {https://doi.org/10.1103/PhysRevLett.115.122301} {\bibfield  {journal}
  {\bibinfo  {journal} {Phys. Rev. Lett.}\ }\textbf {\bibinfo {volume} {115}},\
  \bibinfo {pages} {122301} (\bibinfo {year} {2015})}\BibitemShut {NoStop}%
\bibitem [{\citenamefont {Entem}\ \emph {et~al.}(2017)\citenamefont {Entem},
  \citenamefont {Machleidt},\ and\ \citenamefont
  {Nosyk}}]{Entem2017Phys.Rev.C024004}%
  \BibitemOpen
  \bibfield  {author} {\bibinfo {author} {\bibfnamefont {D.~R.}\ \bibnamefont
  {Entem}}, \bibinfo {author} {\bibfnamefont {R.}~\bibnamefont {Machleidt}},\
  and\ \bibinfo {author} {\bibfnamefont {Y.}~\bibnamefont {Nosyk}},\ }\bibfield
   {title} {\bibinfo {title} {{High-quality two-nucleon potentials up to fifth
  order of the chiral expansion}},\ }\href
  {https://doi.org/10.1103/PhysRevC.96.024004} {\bibfield  {journal} {\bibinfo
  {journal} {Phys. Rev. C}\ }\textbf {\bibinfo {volume} {96}},\ \bibinfo
  {pages} {024004} (\bibinfo {year} {2017})}\BibitemShut {NoStop}%
\bibitem [{\citenamefont {Tjon}(1975)}]{Tjon1975Phys.Lett.B217220}%
  \BibitemOpen
  \bibfield  {author} {\bibinfo {author} {\bibfnamefont {J.}~\bibnamefont
  {Tjon}},\ }\bibfield  {title} {\bibinfo {title} {{Bound states of 4He with
  local interactions}},\ }\href
  {https://doi.org/https://doi.org/10.1016/0370-2693(75)90378-0} {\bibfield
  {journal} {\bibinfo  {journal} {Phys. Lett. B}\ }\textbf {\bibinfo {volume}
  {56}},\ \bibinfo {pages} {217} (\bibinfo {year} {1975})}\BibitemShut
  {NoStop}%
\bibitem [{\citenamefont {Margueron}\ \emph {et~al.}(2018)\citenamefont
  {Margueron}, \citenamefont {Hoffmann~Casali},\ and\ \citenamefont
  {Gulminelli}}]{Margueron2018Phys.Rev.C025805}%
  \BibitemOpen
  \bibfield  {author} {\bibinfo {author} {\bibfnamefont {J.}~\bibnamefont
  {Margueron}}, \bibinfo {author} {\bibfnamefont {R.}~\bibnamefont
  {Hoffmann~Casali}},\ and\ \bibinfo {author} {\bibfnamefont {F.}~\bibnamefont
  {Gulminelli}},\ }\bibfield  {title} {\bibinfo {title} {{Equation of state for
  dense nucleonic matter from metamodeling. I. Foundational aspects}},\ }\href
  {https://doi.org/10.1103/PhysRevC.97.025805} {\bibfield  {journal} {\bibinfo
  {journal} {Phys. Rev. C}\ }\textbf {\bibinfo {volume} {97}},\ \bibinfo
  {pages} {025805} (\bibinfo {year} {2018})}\BibitemShut {NoStop}%
\bibitem [{\citenamefont {Akmal}\ \emph {et~al.}(1998)\citenamefont {Akmal},
  \citenamefont {Pandharipande},\ and\ \citenamefont
  {Ravenhall}}]{Akmal1998Phys.Rev.C18041828}%
  \BibitemOpen
  \bibfield  {author} {\bibinfo {author} {\bibfnamefont {A.}~\bibnamefont
  {Akmal}}, \bibinfo {author} {\bibfnamefont {V.~R.}\ \bibnamefont
  {Pandharipande}},\ and\ \bibinfo {author} {\bibfnamefont {D.~G.}\
  \bibnamefont {Ravenhall}},\ }\bibfield  {title} {\bibinfo {title} {{Equation
  of state of nucleon matter and neutron star structure}},\ }\href
  {https://doi.org/10.1103/PhysRevC.58.1804} {\bibfield  {journal} {\bibinfo
  {journal} {Phys. Rev. C}\ }\textbf {\bibinfo {volume} {58}},\ \bibinfo
  {pages} {1804} (\bibinfo {year} {1998})}\BibitemShut {NoStop}%
\bibitem [{\citenamefont {Li}\ \emph {et~al.}(2006)\citenamefont {Li},
  \citenamefont {Lombardo}, \citenamefont {Schulze}, \citenamefont {Zuo},
  \citenamefont {Chen},\ and\ \citenamefont {Ma}}]{Li2006Phys.Rev.C047304}%
  \BibitemOpen
  \bibfield  {author} {\bibinfo {author} {\bibfnamefont {Z.~H.}\ \bibnamefont
  {Li}}, \bibinfo {author} {\bibfnamefont {U.}~\bibnamefont {Lombardo}},
  \bibinfo {author} {\bibfnamefont {H.-J.}\ \bibnamefont {Schulze}}, \bibinfo
  {author} {\bibfnamefont {W.}~\bibnamefont {Zuo}}, \bibinfo {author}
  {\bibfnamefont {L.~W.}\ \bibnamefont {Chen}},\ and\ \bibinfo {author}
  {\bibfnamefont {H.~R.}\ \bibnamefont {Ma}},\ }\bibfield  {title} {\bibinfo
  {title} {{Nuclear matter saturation point and symmetry energy with modern
  nucleon-nucleon potentials}},\ }\href
  {https://doi.org/10.1103/PhysRevC.74.047304} {\bibfield  {journal} {\bibinfo
  {journal} {Phys. Rev. C}\ }\textbf {\bibinfo {volume} {74}},\ \bibinfo
  {pages} {047304} (\bibinfo {year} {2006})}\BibitemShut {NoStop}%
\bibitem [{\citenamefont {Hu}\ \emph {et~al.}(2017)\citenamefont {Hu},
  \citenamefont {Zhang}, \citenamefont {Epelbaum}, \citenamefont
  {Mei\ss{}ner},\ and\ \citenamefont {Meng}}]{Hu2017Phys.Rev.C034307}%
  \BibitemOpen
  \bibfield  {author} {\bibinfo {author} {\bibfnamefont {J.}~\bibnamefont
  {Hu}}, \bibinfo {author} {\bibfnamefont {Y.}~\bibnamefont {Zhang}}, \bibinfo
  {author} {\bibfnamefont {E.}~\bibnamefont {Epelbaum}}, \bibinfo {author}
  {\bibfnamefont {U.-G.}\ \bibnamefont {Mei\ss{}ner}},\ and\ \bibinfo {author}
  {\bibfnamefont {J.}~\bibnamefont {Meng}},\ }\bibfield  {title} {\bibinfo
  {title} {{Nuclear matter properties with nucleon-nucleon forces up to fifth
  order in the chiral expansion}},\ }\href
  {https://doi.org/10.1103/PhysRevC.96.034307} {\bibfield  {journal} {\bibinfo
  {journal} {Phys. Rev. C}\ }\textbf {\bibinfo {volume} {96}},\ \bibinfo
  {pages} {034307} (\bibinfo {year} {2017})}\BibitemShut {NoStop}%
\bibitem [{\citenamefont {Epelbaum}\ \emph {et~al.}(2009)\citenamefont
  {Epelbaum}, \citenamefont {Hammer},\ and\ \citenamefont
  {Mei\ss{}ner}}]{Epelbaum2009Rev.Mod.Phys.17731825}%
  \BibitemOpen
  \bibfield  {author} {\bibinfo {author} {\bibfnamefont {E.}~\bibnamefont
  {Epelbaum}}, \bibinfo {author} {\bibfnamefont {H.-W.}\ \bibnamefont
  {Hammer}},\ and\ \bibinfo {author} {\bibfnamefont {U.-G.}\ \bibnamefont
  {Mei\ss{}ner}},\ }\bibfield  {title} {\bibinfo {title} {{Modern theory of
  nuclear forces}},\ }\href {https://doi.org/10.1103/RevModPhys.81.1773}
  {\bibfield  {journal} {\bibinfo  {journal} {Rev. Mod. Phys.}\ }\textbf
  {\bibinfo {volume} {81}},\ \bibinfo {pages} {1773} (\bibinfo {year}
  {2009})}\BibitemShut {NoStop}%
\bibitem [{\citenamefont {Machleidt}\ and\ \citenamefont
  {Entem}(2011)}]{Machleidt2011Phys.Rep.175}%
  \BibitemOpen
  \bibfield  {author} {\bibinfo {author} {\bibfnamefont {R.}~\bibnamefont
  {Machleidt}}\ and\ \bibinfo {author} {\bibfnamefont {D.}~\bibnamefont
  {Entem}},\ }\bibfield  {title} {\bibinfo {title} {{Chiral effective field
  theory and nuclear forces}},\ }\href
  {https://doi.org/10.1016/j.physrep.2011.02.001} {\bibfield  {journal}
  {\bibinfo  {journal} {Phys. Rep.}\ }\textbf {\bibinfo {volume} {503}},\
  \bibinfo {pages} {1} (\bibinfo {year} {2011})}\BibitemShut {NoStop}%
\bibitem [{\citenamefont {Hammer}\ \emph {et~al.}(2020)\citenamefont {Hammer},
  \citenamefont {K\"onig},\ and\ \citenamefont {van
  Kolck}}]{Hammer2020Rev.Mod.Phys.025004}%
  \BibitemOpen
  \bibfield  {author} {\bibinfo {author} {\bibfnamefont {H.-W.}\ \bibnamefont
  {Hammer}}, \bibinfo {author} {\bibfnamefont {S.}~\bibnamefont {K\"onig}},\
  and\ \bibinfo {author} {\bibfnamefont {U.}~\bibnamefont {van Kolck}},\
  }\bibfield  {title} {\bibinfo {title} {{Nuclear effective field theory:
  Status and perspectives}},\ }\href
  {https://doi.org/10.1103/RevModPhys.92.025004} {\bibfield  {journal}
  {\bibinfo  {journal} {Rev. Mod. Phys.}\ }\textbf {\bibinfo {volume} {92}},\
  \bibinfo {pages} {025004} (\bibinfo {year} {2020})}\BibitemShut {NoStop}%
\bibitem [{\citenamefont {Epelbaum}\ and\ \citenamefont
  {Gegelia}(2012)}]{Epelbaum2012Phys.Lett.B338344}%
  \BibitemOpen
  \bibfield  {author} {\bibinfo {author} {\bibfnamefont {E.}~\bibnamefont
  {Epelbaum}}\ and\ \bibinfo {author} {\bibfnamefont {J.}~\bibnamefont
  {Gegelia}},\ }\bibfield  {title} {\bibinfo {title} {{Weinberg's approach to
  nucleon-nucleon scattering revisited}},\ }\href
  {https://doi.org/https://doi.org/10.1016/j.physletb.2012.08.025} {\bibfield
  {journal} {\bibinfo  {journal} {Phys. Lett. B}\ }\textbf {\bibinfo {volume}
  {716}},\ \bibinfo {pages} {338} (\bibinfo {year} {2012})}\BibitemShut
  {NoStop}%
\bibitem [{\citenamefont {Ren}\ \emph {et~al.}(2018)\citenamefont {Ren},
  \citenamefont {Li}, \citenamefont {Geng}, \citenamefont {Long}, \citenamefont
  {Ring},\ and\ \citenamefont {Meng}}]{Ren2018Chin.Phys.C014103}%
  \BibitemOpen
  \bibfield  {author} {\bibinfo {author} {\bibfnamefont {X.-L.}\ \bibnamefont
  {Ren}}, \bibinfo {author} {\bibfnamefont {K.-W.}\ \bibnamefont {Li}},
  \bibinfo {author} {\bibfnamefont {L.-S.}\ \bibnamefont {Geng}}, \bibinfo
  {author} {\bibfnamefont {B.}~\bibnamefont {Long}}, \bibinfo {author}
  {\bibfnamefont {P.}~\bibnamefont {Ring}},\ and\ \bibinfo {author}
  {\bibfnamefont {J.}~\bibnamefont {Meng}},\ }\bibfield  {title} {\bibinfo
  {title} {{Leading order relativistic chiral nucleon-nucleon interaction*}},\
  }\href {https://doi.org/10.1088/1674-1137/42/1/014103} {\bibfield  {journal}
  {\bibinfo  {journal} {Chin. Phys. C}\ }\textbf {\bibinfo {volume} {42}},\
  \bibinfo {pages} {014103} (\bibinfo {year} {2018})}\BibitemShut {NoStop}%
\bibitem [{\citenamefont {Ren}\ \emph {et~al.}(2022)\citenamefont {Ren},
  \citenamefont {Epelbaum},\ and\ \citenamefont
  {Gegelia}}]{Ren2022Phys.Rev.C034001}%
  \BibitemOpen
  \bibfield  {author} {\bibinfo {author} {\bibfnamefont {X.-L.}\ \bibnamefont
  {Ren}}, \bibinfo {author} {\bibfnamefont {E.}~\bibnamefont {Epelbaum}},\ and\
  \bibinfo {author} {\bibfnamefont {J.}~\bibnamefont {Gegelia}},\ }\bibfield
  {title} {\bibinfo {title} {{Nucleon-nucleon scattering up to
  next-to-next-to-leading order in manifestly Lorentz-invariant chiral
  effective field theory: Peripheral phases}},\ }\href
  {https://doi.org/10.1103/PhysRevC.106.034001} {\bibfield  {journal} {\bibinfo
   {journal} {Phys. Rev. C}\ }\textbf {\bibinfo {volume} {106}},\ \bibinfo
  {pages} {034001} (\bibinfo {year} {2022})}\BibitemShut {NoStop}%
\bibitem [{\citenamefont {Lu}\ \emph {et~al.}(2022)\citenamefont {Lu},
  \citenamefont {Wang}, \citenamefont {Xiao}, \citenamefont {Geng},
  \citenamefont {Meng},\ and\ \citenamefont
  {Ring}}]{Lu2022Phys.Rev.Lett.142002}%
  \BibitemOpen
  \bibfield  {author} {\bibinfo {author} {\bibfnamefont {J.-X.}\ \bibnamefont
  {Lu}}, \bibinfo {author} {\bibfnamefont {C.-X.}\ \bibnamefont {Wang}},
  \bibinfo {author} {\bibfnamefont {Y.}~\bibnamefont {Xiao}}, \bibinfo {author}
  {\bibfnamefont {L.-S.}\ \bibnamefont {Geng}}, \bibinfo {author}
  {\bibfnamefont {J.}~\bibnamefont {Meng}},\ and\ \bibinfo {author}
  {\bibfnamefont {P.}~\bibnamefont {Ring}},\ }\bibfield  {title} {\bibinfo
  {title} {{Accurate Relativistic Chiral Nucleon-Nucleon Interaction up to
  Next-to-Next-to-Leading Order}},\ }\href
  {https://doi.org/10.1103/PhysRevLett.128.142002} {\bibfield  {journal}
  {\bibinfo  {journal} {Phys. Rev. Lett.}\ }\textbf {\bibinfo {volume} {128}},\
  \bibinfo {pages} {142002} (\bibinfo {year} {2022})}\BibitemShut {NoStop}%
\end{thebibliography}

%

\end{document}